\documentclass[useAMS,usenatbib]{mn2e}
\usepackage{amssymb,amsmath}
\usepackage{graphicx,epsfig,rotating}
\title[Angular and spatial clustering of quasar candidates from SDSS NBCKDE]{Angular and spatial clustering of photometrically\\ classified quasar candidates from SDSS NBCKDE}
\author[G.~Ivashchenko, O.~Vasylenko, A.\,V.~Tugay]{Ivashchenko~G.$^{1}$\thanks{E-mail: g.ivashchenko@gmail.com}, Vasylenko~O.$^{2}$\thanks{E-mail: OlyaVasilenko@bigmir.net}, Tugay~A.~V.$^{2}$\\
$^1$Astronomical Observatory of the Taras Shevchenko National University of Kyiv, Observatorna str., 3, 04058, Kyiv, Ukraine\\
$^2$Faculty of Physics of the Taras Shevchenko National University of Kyiv, Glushkova ave., 4, 03127, Kyiv, Ukraine\\
}
\begin{document}
\date{Accepted \underline{\hspace*{2cm}}. Received \underline{\hspace*{2cm}}; in original form \underline{\hspace*{2cm}}}
\pagerange{\pageref{firstpage}--\pageref{lastpage}} \pubyear{0000}
\maketitle

\label{firstpage}

\begin{abstract}
Quasar distribution is our main source of information about the large-scale distribution of matter at redshifts greater than 1. It can also shed light on the open questions of quasar formation. The present paper analyses the quasar clustering using the two-point correlation function (2pCF) and the largest existing sample of photometrically selected quasars: the SDSS NBCKDE catalogue (from the SDSS DR6). A new technique of random catalogue generation was developed for this purpose, that allows to take into account the original homogeneity of the survey without knowledge of its imaging mask. When averaged over photometrical redshifts $0.8<z_{phot}<2.2$ the 2pCF of photometrically selected quasars is found to be approximated well with the power law $w(\theta)=\left(\theta/\theta_{0}\right)^{-\alpha}$ with $\theta_{0}=4''.5\pm1''.4$ and $\alpha=0.94\pm0.06$ over the range $1'<\theta<40'$. It agrees well with previous results by \citet{myers_2006,myers_2007}, obtained for samples of NBCKDE quasars with similar 
mean redshift, but averaged over broader $z_{phot}$ range. The parameters of the deprojected 2pCF averaged over the same redshift range and modelled with a power law $\xi(r)=\left(r/r_{0}\right)^{-\gamma}$, are $r_{0}=7.81^{+1.18}_{-1.16}\,h^{-1}$\,Mpc, $\gamma=1.94\pm0.06$, which are in perfect agreement with previous results from spectroscopic surveys. We confirm the evidence for an increase of the clustering amplitude with redshift, and find no evidence for luminosity dependence of the quasar clustering. The latter is consistent with the models of the quasar formation, in which bright and faint quasars are assumed to be similar sources, hosted by dark matter halos of similar masses, but observed at different stages of their evolution. Comparison of our results with studies of the X-ray selected AGNs with similar redshift shows that the clustering amplitude of optically selected quasars is similar to that of X-ray selected quasars, but lower than that of samples of all X-ray selected AGNs. As the samples 
of all X-ray selected AGNs contain AGNs of both types, our result serves as an evidence for different types of AGNs to reside in different environments.
\end{abstract}

\begin{keywords}
cosmology: observations -- large-scale structure, surveys
\end{keywords}

\section{Introduction}\label{sec:1}

\indent\indent Quasars, the brightest extragalactic objects, play an important role in study of the large-scale structure of the Universe. One of important characteristics of matter spatial inhomogeneity, that may be compared to cosmological theories of structure formation, is the two-point correlation function  (2pCF; e.\,g. \citealt{peebles_book}) of quasars, $\xi(r)$. The main results on quasars' clustering at $z<2.2$ (see, e.\,g., \citealt{croom_2005,myers_2006,myers_2007,ross_2009} and references therein) were obtained using the 2-degree Field (2dF) Quasar Survey\footnote{http://www.2dfquasar.org} (2QZ; \citealt{Croom_2004}) and the Sloan Digital Sky Survey (SDSS)\footnote{http://www.sdss.org}, the second stage of which has been completed with the 7th release \citep{Abazadjian_2009}. The first results on quasar clustering for more 
complete sample of quasars with $z>2.2$ \citep{white+2012} appeared recently by virtue of the 9th data release of the SDSS \citep{dr9} which comprises the Baryon Oscillation Spectroscopic Survey (BOSS).

The study of the matter distribution with the help of extragalactic redshift surveys faces two main problems. Firstly, galaxy and quasar surveys give us information only about the distribution of luminous matter which is biased relative to dark matter \citep{kaiser_1984}. This bias is expected to be non-linear, scale-dependent and stochastic \citep{mo+1996,sheth+1999}, but the simplest, linear, scale-independent bias parameter, which relates the 2pCFs of the dark+luminous matter, $\xi_{m}(r)$, and that of the luminous matter, e.\,g. quasars,  $\xi(r)$, as
\begin{equation}\label{eq:bias}
 \xi(r)=b^{2}\xi_{m}(r),
\end{equation}
has proved to be an extremely useful first order approximation \citep{kaiser_1984,bardeen+1986,cole+1989}. This parameter depends on the nature of extragalactic objects, i.e., on morphological type \citep{einasto_2007,ross_2007}, star formation rate \citep{owners_2007}, color-index \citep{coil_2008}, or the luminosity \citep{beisbart_2000,zehavi_2005,sorrentino_2006,coil_2008}. The latter dependence is explained within the framework of the hierarchical model of the structure formation, in which more massive, and therefore more luminous galaxies, have to reside in more massive dark matter halos, and thus to be more clustered. It is unclear however whether this relation can be applied to quasar clustering, because of obviously more complicated process of the quasar formation, which in its turn still does not have a concordance model. Several models of quasar formation were presented in the literature (e.\,g. \citealt{kauffmann+2000,hopkins+2005,hopkins+2008,croton+2006,thacker+2008}) and based on study of the 
luminosity dependence of the active galactic nuclei (AGN) clustering, some of them were completely ruled out, while some models are partially supported by observations so far. Hence study of the luminosity dependence of the quasar clustering in different wavelength bands can help to clarify not only the relation between the luminous and dark matter clustering, but also the physics of the quasar formation. It was also shown by several authors (see e.\,g. \citealt{croom_2005,weinstein_2004,porciani_2004,myers_2006,myers_2007,daAngela_2008,Mountrichas_2009}) that the linear bias parameter of quasars evolves with redshift. 

Secondly, analysis of the distribution of extragalactic objects is related to the observed redshifts of these objects. These redshifts are the only tool for measuring distance between the objects on cosmological scales, but they are `contaminated' by measurement errors and non-Hubble motions. It results in the so called redshift-space distortions, namely \citet{kaiser_1987} and `Finger of God' effects (see, e.\,g., \citealt{daAngela_2008,Mountrichas_2009,ross_2009,ivashchenko_2010} and references therein). 

To avoid the problem of the redshift-space distortion the projected, spatial $w(\sigma)$ and angular $w(\theta)$, 2pCF are usually used. In the first case the projections $\sigma$ of the 3-dimensional distances $r$ on the plane perpendicular to the line of sight are used, and the parameters of the 3-dimensional 2pCF $\xi(r)$ can be reconstructed from the parameters of $w(\sigma)$. Here the assumption, that the influence of the redshift inexactness causes a negligible effect on $\sigma$ compared to $r$, is usually made. In the second case only the angular distances on the sky plane are used, and the parameters of $\xi(r)$ can be reconstructed from $w(\theta)$ using Limber's equation \citep{limber_1953}. This process of deprojection of the 2pCF is more complicated than in the first case because it comprises the usage of the redshift distribution function of objects and analytical functions describing the clustering evolution. Nevertheless, taking into account smaller sizes and larger volumes of quasar 
spectroscopic surveys compared to  surveys of galaxies, the angular 2pCF is still appealing for quasars because it does not include redshift information and thus allows the use of catalogues of photometrically classified quasars, which probe to deeper flux limits, and so typically contain an order of magnitude larger number of objects than spectroscopic quasar catalogues.  

Study of the angular clustering of photometrically selected quasars was pioneered by \citet{myers_2006,myers_2007,myers_2007_2} using the catalogues of NBCKDE photometrically selected quasar candidates from SDSS \citep{richards_2004,richards_2009} based on the Early and the 4th data releases of SDSS. It was shown that the angular 2pCF of quasars is fitted well with a power law  $w(\theta)=\left(\theta/\theta_{0}\right)^{-\alpha}$, where $\alpha$ is a slope of the angular 2pCF. Myers et al. noted that the angular 2pCF does not evolve strongly with redshift, suggesting that the bias of quasars relative to underlying dark matter climbs steeply out to redshift 2.2. The reconstructed parameters of the 3D clustering by these authors agree well with the `direct' measurements of 3D clustering of quasars conducted e.\,g. by \citet{porciani_2004}, \citet{croom_2005}, \citet{daAngela_2008}. It is interesting to compare the  clustering of quasars with those of X-ray point-like sources, a large majority of which are 
supposed to be AGN. The study of the X-ray AGN clustering has been conducted intensively since the release of the X-ray sky surveys, such as ROSAT All-Sky Survey, XMM-Newton and Chandra, and especially after the cross-correlation of X-ray surveys with optical redshift surveys of AGNs (for some latest results see e.\,g. \citealt{Allevato+2011,elyiv_2011}). 

In this paper we present our study of the angular and spatial clustering of photometrically selected quasar candidates from the SDSS NBCKDE catalogue \citep{richards_2009}, based on the 6th release of the SDSS. This catalogue is 3 times larger than that used in \citealt{myers_2007,myers_2007_2}, which allowed us to place more tight restrictions on the samples, namely to reject objects with photometric redshift determination probability less than 0.5, and to construct the main sample only from objects with redshifts within the `SDSS window', unlike \citet{myers_2007}, who used the whole redshift range of the catalogue. This or slightly wider `window' is traditionally used in studies of spectroscopic samples because of the better efficiency of the photometric selection algorithm of quasars, and because of the peak of the quasar distribution within this redshift range. Although we follow \citet{myers_2006} in deprojection of the angular 2pCF and calculation of the bias parameter, our algorithm of the random 
catalogue generation differs from the traditional technique, used by \citet{myers_2006}, which is an important issue in verification of the 2pCF results.

The samples, procedure their selection and properties, along with a random catalogue generation method are described in Sec.\,\ref{sec:2}. The technique of the angular 2pCF calculation, its results and discussion are presented in Sec.\,\ref{sec:3}. Reconstruction of the spatial clustering parameters along with the corresponding technique are described in Sec.\,\ref{sec:4}. Sec.\,\ref{sec:5} contains the results on the luminosity and redshift dependence of the quasar clustering along with comparison with other studies of optical quasar samples and samples of X-ray selected AGNs. Finally, in Sec.~\ref{sec:6} we sum up the results.  Throughout the paper we calculate distances within the spatially flat $\Lambda$CDM model with $\Omega_{M}=0.28$ and $H_{0}=71$\,km\,s$^{-1}$\,Mpc$^{-1}$.

\vspace*{-3ex}
\section{Data}\label{sec:2}
\subsection{SDSS NBCKDE catalogue}\label{sec:2-1}

\indent\indent Our sample is taken from the SDSS NBCKDE Catalogue of Photometrically Classified Quasar Candidates \citep{richards_2009} that contains  1\,015\,082 quasar candidates selected from the photometric imaging data of the SDSS using a non-parametric Bayesian classification kernel density estimator (NBC-KDE). The objects are all point sources to a limiting magnitude of $i = 21.3$ derived from 8417~deg$^{2}$ of imaging from the SDSS Data Release 6. The authors have demonstrated that the overall efficiency (quasars/quasar candidates) of the catalogue is 95\% (e.\,g., \citealt{richards_2004,myers_2007,richards_2009}).

\subsection{The full, high- and low-reddening samples}\label{sec:2-2}

\begin{figure*}
\centering
\begin{minipage}{.90\textwidth}
\centering
\epsfig{figure=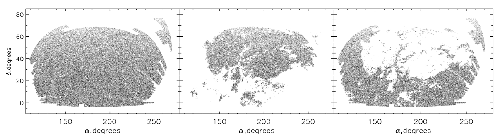,width=.99\linewidth}\\
\epsfig{figure=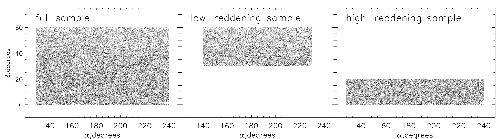,width=.99\textwidth}
\caption{Top (left to right): sky coverage of all selected objects and objects with low ($E(B-V)\leq0.0217$) and high ($E(B-V)>0.0217$) reddening. Bottom: sky coverage of the \textit{full}, \textit{low-} and \textit{high-reddening} samples, described in Sec.\,\ref{sec:2-2}.}\label{fig:ad-full-flh}
\end{minipage}
\end{figure*}

\indent\indent For our study we selected only the objects with photometric redshifts within the range $0.8\leq{z_{phot}}\leq2.2$ and photometric redshift range probability $z_{photprob}>0.5$. This redshift range is known as the `SDSS window' \citep{weinstein_2004}, i.\,e. the redshift range where the algorithm for photometric selection of quasar candidates is the most efficient. As the sky coverage of SDSS contains one big `piece' and three narrow near-equatorial `stripes', we excluded these stripes to reduce boundary effects, thus our right ascension range is $100^{\circ}\leq\alpha\leq270^{\circ}$. The resulting (\textit{initial}) sample contains 320\,761 objects. Its sky coverage is presented in Fig.~\ref{fig:ad-full-flh} (top left panel). 

As the main criterion of the KDE photometric selection algorithm is based on quasar magnitudes and colours, that have been corrected for Galactic extinction using the maps of \citet{schlegel_1998}, any systematic errors in the reddening model can induce additional effects on clustering results (see \citet{myers_2006,ross_2009} for a discussion of these effects). That is why following \citet{ross_2009,ivashchenko_2010} we divided our sample into low- ($E(B-V)\leq0.0217$) and high-reddening ($E(B-V)>0.0217$) parts. The numbers of objects in these parts are 128\,757 and 192\,004 for low and high reddening respectively, and their sky coverage is shown in Fig.~\ref{fig:ad-full-flh} (top middle and right panels). 

\begin{figure*}
\centering
\begin{minipage}{.90\linewidth}
\centering
\epsfig{figure=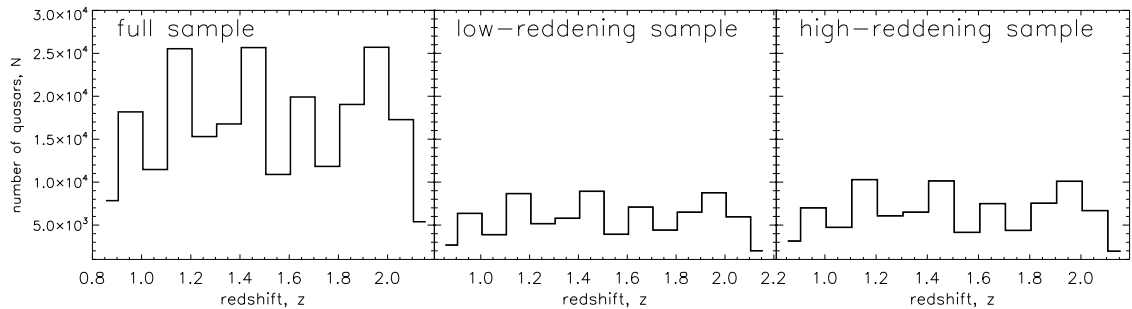,width=.95\textwidth}
\caption{Redshift distribution of the \textit{full}, \textit{low-} and \textit{high-reddening}  samples, described in Sec.\,\ref{sec:2-2}.}\label{fig:z-flh}
\end{minipage}
\end{figure*}

Our technique of random catalogues generation (see Sec.~\ref{sec:2-6}) requires as smooth as possible shape of the sky coverage. Therefore to study the angular 2pCF and possible influence of the different reddening on the results we did not the subsamples described above, but three other samples selected from the \textit{initial} sample. The first, \textit{full}, sample is a patch of the sky with $130^{\circ}<\alpha<240^{\circ}$, $0^{\circ}<\delta<60^{\circ}$, containing 230\,829 objects. The second one is a patch of the sky with $140^{\circ}<\alpha<230^{\circ}$, $30^{\circ}<\delta<60^{\circ}$, containing 80\,107 objects (71\,190, or 89\%, of which are `low-reddening' quasars according to the definition given above) and the third one is a patch of the sky with $130^{\circ}<\alpha<240^{\circ}$, $0^{\circ}<\delta<20^{\circ}$, containing 90\,185 objects (79\,450, or 88\%, of which are `high-reddening' quasars). We formally call these last two samples \textit{low-} and \textit{high-reddening} samples 
respectively. Their sky coverage and the redshift distribution are shown in the bottom panels of Fig.~\ref{fig:ad-full-flh} and in Fig.~\ref{fig:z-flh}. One can see, that the redshift distribution for all three samples are quite similar.

Note that our various cuts, coupled with the small increase in imaging coverage between SDSS DR4 and SDSS DR6, mean that our sample is about the same size as the sample of 299\,276 photometrically classified quasars used for similar studies in \citet{myers_2007_2}. Thus, our work should be viewed as a comprehensive check on the techniques and results of \citet{myers_2007_2}.

\subsection{Contamination of the sample}\label{sec:2-3}

\indent\indent The KDE photometric selection algorithm for quasar candidates has its limitations. Based on classifying simulated quasars, \citet{richards_2004} found that the KDE technique is 95\% efficient, with 5\% stellar contamination. To check for additional sources of contamination (such as those already noted in \citealt{myers_2007_2}) we selected at random 30\,000 objects from the \textit{initial} sample (about 10\% of the sample) and examined them by eye using the SDSS SkyServer web-service\footnote{http://cas.sdss.org/}. Among these 30,000 objects, 28 appeared to be bright blue parts of spiral galaxies, probably star formation regions, 1 is a faint extended object, probably a marginally resolved galaxy or nebula (`STAR' according to SDSS nomenclature), and 4 lie close to diffraction spikes around bright stars. Thus we can conclude, that an additional contamination of the sample by misclassified point-like sources is only $\sim0.1$\%, which can be neglected in comparison with the stellar 
contamination. These results are very similar to the study of \citet{myers_2007_2}. We note that \citet{myers_2007_2} pointed out that contaminating objects such as star-forming regions in nearby galaxies are highly correlated on scales of a few-tenths of an arcsecond. Therefore, our clustering measurements may be significantly inflated on small scales.

\subsection{Samples with different luminosities}\label{sec:2-4}

\indent\indent To check the luminosity dependence of the quasar clustering we selected three samples with different luminosity. As a characteristic of luminosity, the absolute magnitude in $i$-band, $M_{i}$, was chosen. Its value for each object was calculated using the Pogson's formula with corrections in the following form:
\begin{equation*}
 M_{i}=m_{i}-25-5\log\left(d_{L}\right)+e_{i}+K(z),
\end{equation*}
where $d_{L}$ is the luminosity distance of the object in Mpc, $e_{i}$ is the galactic extinction in $i$-band, and $K(z)$ is the K-correction. The values of $e_{i}$, obtained from the maps by \citet{schlegel_1998}, were taken from the catalogue. The values of $K(z)$ were obtained from \citet{richards_2006}. 

To separate the dependencies on luminosity and redshift, but preserve the sample as large as possible, we restricted the redshift range for the samples with different $M_{i}$ to $0.8\leqslant{z}\leqslant2.2$, and selected three subsamples with approximately equal number of objects. These subsamples, which we called \textit{bright}, \textit{medium-M} and \textit{faint} samples, contain 73\,307, 77\,891 and 73\,262 objects with absolute magnitudes in $i$-band within the ranges $-27.0\leq M_{i}\leq-25.3$, $-25.3\leq M_{i}\leq-24.5$, and $-24.5\leq M_{i}\leq-22.5$. Corresponding mean values of $M_{i}$ are $-25.89$, $-24.90$, and $-23.33$. Note, that we used photometric redshifts to calculate the absolute magnitudes, therefore this division is only approximate.

\begin{figure*}
\centering
\begin{minipage}{.90\linewidth}
\centering
\epsfig{figure=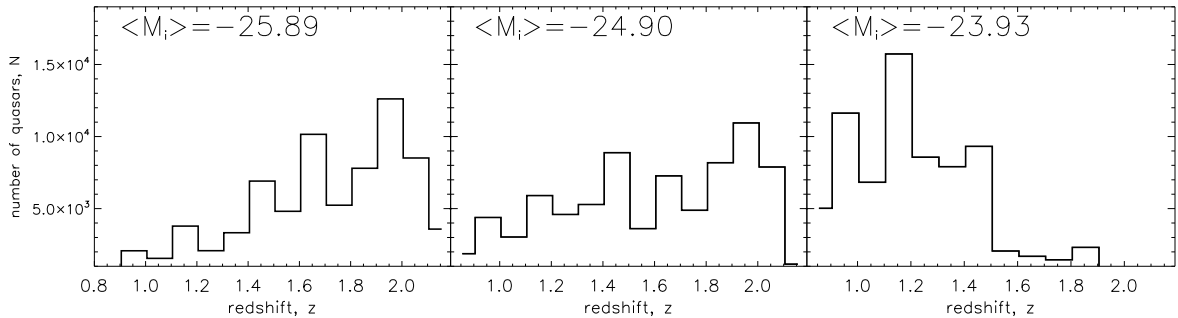,width=.99\textwidth}
\caption{Redshift distribution of (left to right) \textit{bright}, \textit{medium} and \textit{faint} quasars.}\label{fig:z-lum}
\end{minipage}
\end{figure*}

The redshift distributions of these subsamples are presented in Fig.~\ref{fig:z-lum}. It is clearly seen, that even when choosing the relatively narrow $z$-range, the distribution of the faintest objects differs from the distribution of other objects and has a peak at lower redshifts. It means, that any dependence on luminosity, which can be obtained, would be `contaminated' by the redshift-dependence.

\subsection{Samples with different redshifts}\label{sec:2-5}

\begin{figure*}
\centering
\begin{minipage}{.90\linewidth}
\centering
\epsfig{figure=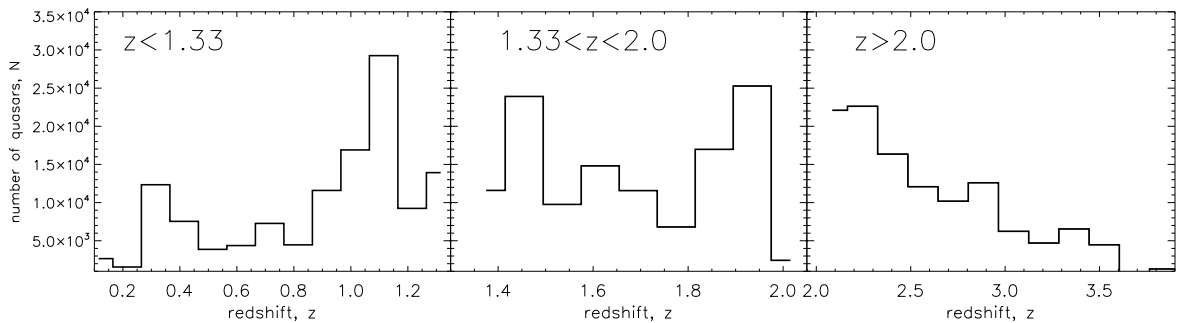,width=.99\textwidth}
\caption{Redshift distribution of objects in samples of (left to right) \textit{nearby}, \textit{medium-z} and \textit{distant} quasars.}\label{fig:z-evo}
\end{minipage}
\end{figure*}

\indent\indent To study the redshift dependence of the quasar clustering, i.\,e. the clustering evolution, we chose all the objects from the SDSS NBCKDE catalogue with the same $\alpha$, $\delta$, and $z_{prob}$ restrictions as for the \textit{full} sample, but with the whole redshift range, and divided it into three parts with approximately equivalent number of objects in each. The number of objects in the obtained samples with redshift ranges $z<1.33$, $1.33<z<2.0$, and $z>2.0$ (we called them the {\it nearby}, {\it medium-z}, and {\it distant} samples) are 125\,021, 123\,142 and 122\,057. Their redshift distributions are presented in Fig.~\ref{fig:z-evo}.

The main characteristics of all considered samples are presented in Table~\ref{tab:samples}: $\alpha$ and $\delta$ ranges, redshift range, the mean redshift, $\bar{z}$, and the number of objects, $N$. 

\begin{table}
 \centering
\caption{Characteristics of the samples used in the present work. See explanation in the text (Sec.\,\ref{sec:2}).}\label{tab:samples}
\vspace*{1ex}
\begin{tabular}{|c|c|c|c|c|c|}
 \hline
sample  & $\alpha$, deg & $\delta$, deg & $z$ range & $N$ & $\bar{z}$ \\
 \hline
{\it full} & $130..240$ & $0..60$ &  & $230 829$ & $1.51$ \\
{\it low-red.} & $140..230$ & $30..60$ & $0.8-2.2$ & $80 107$ & $1.51$ \\
{\it high-red.} & $130..240$ & $0..20$ &  & $90 185$ & $1.50$ \\
 \hline
{\it bright} & & & & $73307$ & $1.69$ \\
{\it medium-M} & $130..240$ & $0..60$ & $0.8-2.2$ & $77891$ & $1.58$  \\
{\it faint} & & & & $73262$ & $1.23$ \\
 \hline
{\it nearby} & & & $<1.33$ & $125021$ & $0.89$ \\
{\it medium-z} & $130..240$ & $0..60$ & $1.33-2.0$ & $123142$ & $1.68$ \\
{\it distant} & & & $>2.0$ & $122057$ & $2.63$ \\
 \hline
\end{tabular}
\end{table}

\subsection{Random samples}\label{sec:2-6}

\indent\indent The random catalogues were generated with the help of the following technique, the idea of which was proposed by \citet{ivashchenko_2008_lum}. Firstly the sky area covered by the sample was divided into `square' cells and then filled with the same number of random points (random $\alpha$ and $\delta$) as in the initial sample with homogeneous distribution along $\alpha$ and $\cos\delta$-distribution along $\delta$. By `squared' we mean a quadrangle with similar number of degrees in $\alpha$ and $\delta$ when projected in Equatorial Coordinates. This technique has the same idea of preserving the original inhomogeneity of the sky coverage by the sample as the usage of the imaging mask (see e.\,g. \citealt{myers_2006}), but unlike that technique, changes in density due to conditions of observations should be naturally mimicked in the random catalogue. An important aspect of our technique lies in the choice of cell size. Cells should be small enough to reproduce all the sample density fluctuations 
and large enough not to smooth the physical clustering of objects. 

\begin{figure}
\centering
\begin{minipage}{.45\textwidth}
\centering
\epsfig{file=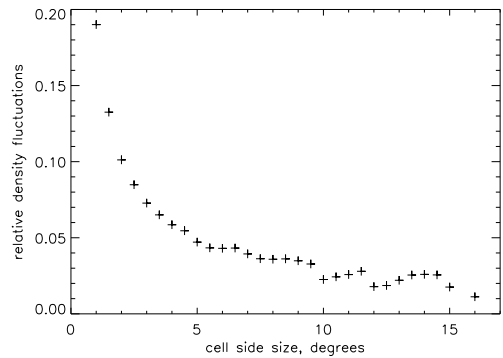, width=0.9\linewidth}
\caption{Relative density fluctuations as a function of the cell side size.}\label{fig:test}
\end{minipage}
\end{figure}

To choose the appropriate cell size we checked different possible sizes from  $1^{\circ}\times1^{\circ}$ up to about $17^{\circ}\times17^{\circ}$. In Fig.~\ref{fig:test} the relative fluctuations (rms) of the number density from cell to cell as a function of the cell side size is presented. As one can see, fluctuations grow with decreasing cell size, as one might expect due to clustering. On some scales about $10^{\circ}$ these fluctuations become constant, which means that these scales are larger than the scales of inhomogeneity. Fluctuations on the largest scales are the result of the small number of cells with these sizes. Hence we choose the smallest possible size to be $10^{\circ}\times10^{\circ}$. We also calculated the angular 2pCF of the {\it full} sample with using different cell sizes and found, that the result does not depend on the cell size for $5^{\circ}-10^{\circ}$ cell size range, but differs for smaller and larger sizes.

\vspace*{-3ex}
\section{Angular 2pCF}\label{sec:3}
\subsection{Technique}\label{sec:3-1}

\indent\indent According to \citet{peebles_book} the angular 2pCF, $w(\theta)$, of a sample of objects determines the probability $dP_1$ to find a neighbour for any object from the sample inside a small solid angle $\delta\Omega$ on the unit sphere at the angular distance $\theta$ as
\begin{equation}\label{eq:2cf-def}
dP_{1} = n_{1}\left[1+w(\theta)\right]\delta\Omega,
\end{equation}
where $n_{1}$ is the number surface density of objects for a given sample. In practice the angular 2pCF of objects is calculated using the numbers of pairs of objects with different separations. In the present work the Landy-Szalay estimator \citep{landy_szalay_1993} was used:
\begin{equation}\label{eq:ls}
 w_{LS}(\theta)=\frac{DD(\theta)}{RR(\theta)}-2\frac{DR(\theta)}{RR(\theta)}+1.
\end{equation}
Here $DD(\theta)$ and $RR(\theta)$ are the numbers of pairs with separations $\theta$ in initial (data-data) and random (random-random) samples, respectively, and  $DR(\theta)$ is a number of cross-pairs between data and random samples (data-random). Using the technique described above we generated 20 random catalogues for each of our samples. Thus in each case the values of $RR$ and $DR$ were calculated as the arithmetic means of 20 corresponding values.

It is known that within some limited scales the angular 2pCF of quasars is fitted well with the power law:
\begin{equation}\label{eq:ang_dkf}
 w(\theta)=\left(\frac{\theta}{\theta_{0}}\right)^{-\alpha}=A\theta^{-\alpha}, 
\end{equation}
where $\theta_{0}$ is the angular correlation length, $\alpha$ is the slope. 

We assume that the main contribution to statistical errors of $w(\theta)$ is made by dispersion of $DD$ values, because usage of 20 random samples for calculation of $DR$ and $RR$ are formally equal to usage of 20 times larger samples. Thus we calculate $\sigma_{w}$ in the following way:
\begin{equation*}\label{eq:err}
 \sigma_{w}^{2}=\left(\frac{\partial{w}}{\partial{DD}}\right)^{2}\sigma_{DD}^{2}=\left(\frac{1}{RR}\right)^{2}\sigma_{DD}^{2},
\end{equation*}
where we used Poisson errors $\sigma_{DD}=\sqrt{DD}$ of $DD$. Note that many authors \citep{shanks_boyle_1994,croom_shanks_1996,Croom_2004,myers_2006} have pointed out that Poisson errors are not valid on scales where pairs become non-independent because the quasars appear in multiple bins of the correlation function at different scales. Therefore our Poisson assumption should underestimate the size of the errors on some scales.

The values of the angular 2pCF parameters along with their extremal errors were estimated from the likelihood function $\mathfrak{L}\sim\exp\left(-\chi^{2}/2\right)$, where 
\begin{equation*}
 \chi^{2}(\theta_{0},\alpha) = \sum\limits_{k}W_{k}\left[w_{k}-\left(\dfrac{\theta_{k}}{\theta_{0}}\right)^{-\alpha}\right]^{2},
\end{equation*}
with normalized weights $W\sim\sigma_{w}^{-2}$.

\subsection{Results and discussion}\label{sec:3-2}

\begin{figure}
\centering
\begin{minipage}{.47\textwidth}
\centering
\epsfig{file=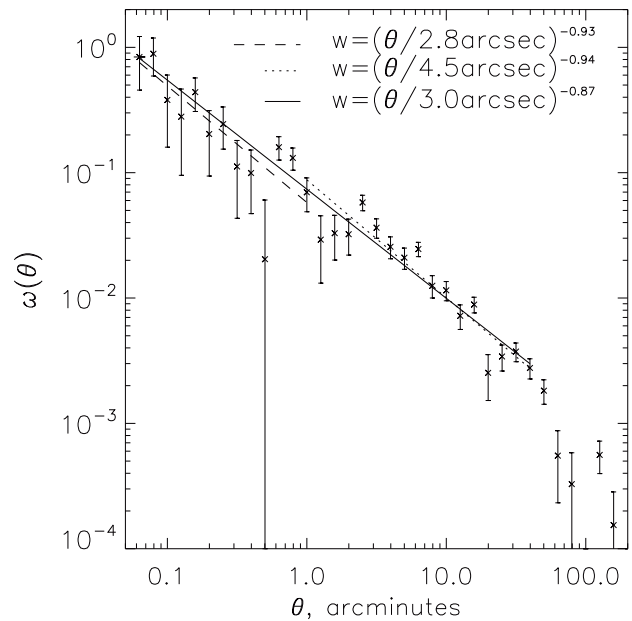, width=0.99\linewidth}
\caption{The angular 2pCF for the \textit{full} sample along with the best fits within the whole range $0.06'-40'$ and separately for ranges $<1'$ and $>1'$.}\label{fig:ls-estim}
\end{minipage}
\end{figure}
\begin{figure}
\centering
\begin{minipage}{.45\textwidth}
\centering
\epsfig{file=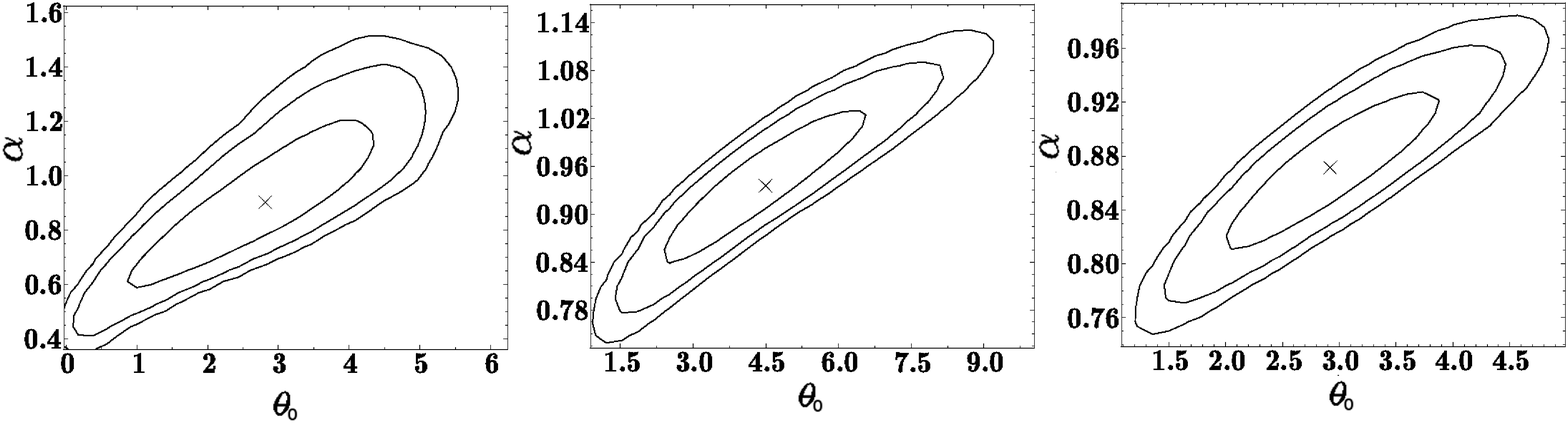, width=0.99\textwidth}
\caption{1,\,2,\,3$\sigma$ confidence levels for $\alpha$ and $\theta_{0}$ parameters of the 2pCF of the \textit{full} sample within (left to right) $0.06'-1'$, $1'-40'$ and $0.06'-40'$. }\label{fig:ls-estim-levels}
\end{minipage}
\end{figure}

\begin{figure}
\centering
\begin{minipage}{.47\textwidth}
\centering
\epsfig{file=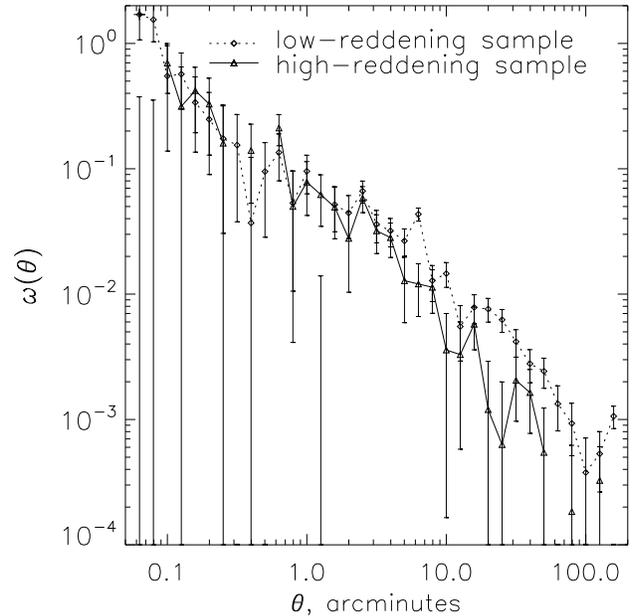, width=0.99\linewidth}
\caption{The angular 2pCF for the \textit{low-} and \textit{high-reddening} samples.}\label{fig:acf-red}
\end{minipage}
\end{figure}

\begin{figure}
\centering
\begin{minipage}{.45\textwidth}
\centering
\epsfig{file=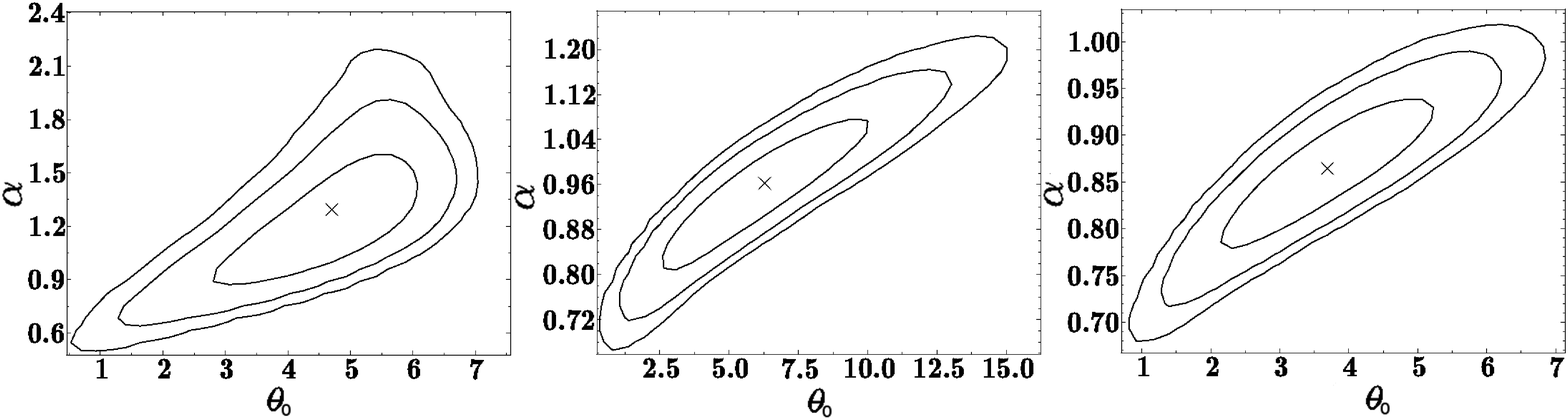, width=0.99\textwidth}
\epsfig{file=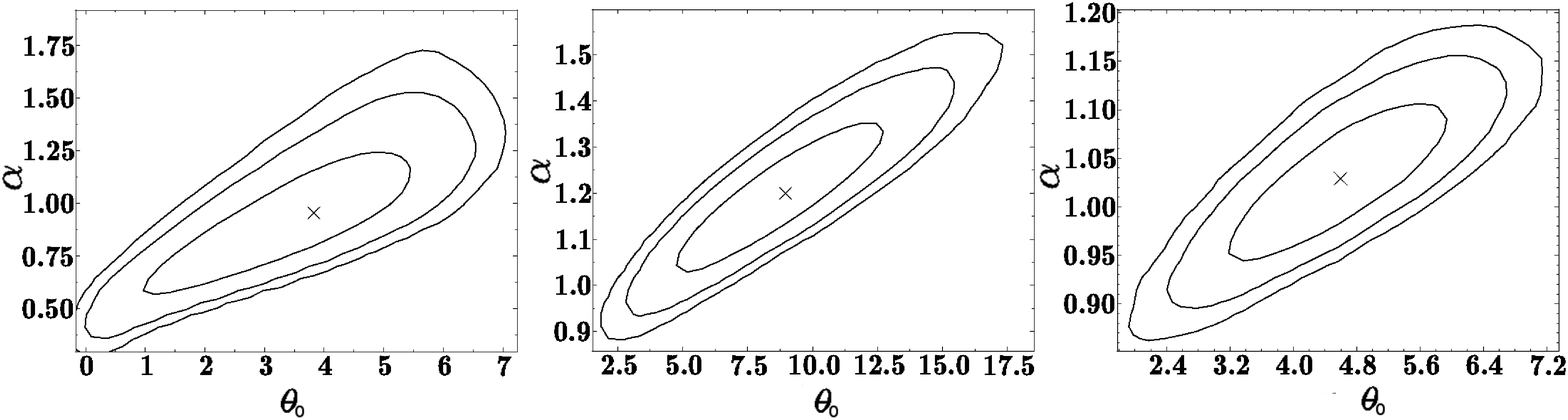, width=0.99\textwidth}
\caption{1,\,2,\,3$\sigma$ confidence levels for $\alpha$ and $\theta_{0}$ parameters of the angular 2pCF for the \textit{low-} (top) and \textit{high-reddening} (bottom) samples within (left to right) $0.06'-1'$, $1'-40'$ and $0.06'-40'$.}\label{fig:acf-red-levels}
\end{minipage}
\end{figure}

\indent\indent In Fig.~\ref{fig:ls-estim} the 2pCF for the \textit{full} sample along with the best fits (within the whole range $0.06'-40'$ and separately for ranges $<1'$ and $>1'$) are shown. The corresponding 1,\,2, and 3$\sigma$ levels of the likelihood function for these fits on the plane $\{\theta_{0},\alpha\}$ are shown in Fig.~\ref{fig:ls-estim-levels}. In Fig.~\ref{fig:acf-red} the 2pCF for the \textit{low-} and \textit{high-reddening} samples are presented, and the corresponding 1,\,2, and 3$\sigma$ levels of the likelihood function for the best fits for the same parameters and the same angular ranges are shown in Fig.~\ref{fig:acf-red-levels}. The best fit values of $\alpha$ and $\theta_{0}$ for all three samples are summarised in Table~\ref{tab:results}. 

As one can see the values of the slopes for the three subsamples agree within 3$\sigma$. Thus one can neglect possible selection effects caused by different Galactic extinction in different parts of the survey within our assumed Poisson errors. Moreover the redshift distribution of quasars (as it is shown in Fig.~\ref{fig:z-flh}) are similar for different reddening. 

Our results agree well with the analyses of \citet{myers_2007,myers_2007_2} on the same angular scales and over a similar redshift range (they obtained $\alpha=0.928\pm0.055$ for the mean redshift $\bar{z}=1.4$). This is unsurprising, as we use slightly smaller samples of the same photometrically classified quasars as used in \citealt{myers_2007,myers_2007_2}, but it is reassuring that our different approaches produce similar results. 

 \begin{table*}
  \centering
  \caption{The slopes $\alpha$ and correlation lengths $\theta_{0}$ for the \textit{full}, \textit{low-reddening} and \textit{high-reddening} samples within different angular ranges.}\label{tab:results}
  \vspace*{1ex}
\begin{tabular}{|c|c|c|c|c|c|c|c|c|c|}
 \hline 
 & \multicolumn{3}{|c|}{\textit{full} sample} & \multicolumn{3}{|c|}{\textit{low-reddening}} & \multicolumn{3}{|c|}{\textit{high-reddening}} \\  
 \cline{2-10}
 &  $\alpha$ & $\theta_{0}$, arcsec & $\chi^{2}/$d.o.f. &  $\alpha$ & $\theta_{0}$, arcsec & $\chi^{2}/$d.o.f. & $\alpha$ & $\theta_{0}$, arcsec & $\chi^{2}/$d.o.f. \\
 \hline
$0.06'-1'$  & $0.93\pm0.20$ & $2.8\pm1.1$ & 1.9  & $0.95^{+0.21}_{-0.22}$ & $3.5\pm1.4$ & 0.8 & $1.24\pm0.24$ & $4.6\pm1.0$ & 0.8\\
$1'-40'$  & $0.94\pm0.06$ & $4.5\pm1.4$ & 2.9 & $1.20^{+0.11}_{-0.10}$ & $9.0\pm2.6$ & 0.9 & $0.96^{+0.08}_{-0.09}$ & $6.4\pm2.5$ & 1.8\\
$0.06'-40'$  & $0.87\pm0.04$ & $3.0\pm0.6$ & 2.3 & $1.03\pm0.05$ & $4.6\pm0.9$ & 0.9 & $0.86^{+0.06}_{-0.05}$ & $3.7\pm1.0$ & 2.9\\
 \hline
  \end{tabular} 
\end{table*}

If quasars are fuelled by galaxy mergers, then there should be an excess of the quasar clustering on small scales. \citet{coil_2007}, studying the cross-correlation of spectroscopically identified quasars from SDSS and DEEP2 with DEEP2 galaxies on scales $0.1\,h^{-1}<\sigma<10\,h^{-1}$\,Mpc, found no significant scale dependence of the quasar clustering at $0.7<z<1.4$, but noted that their quasar sample is too small to provide clustering measures with reasonable errors on small scales. Using 218 quasar pairs with spectroscopic redshifts from the SDSS \citet{hennawi_2006} noted the departure of the 2pCF of quasars with $0.7<z<3$ from a single power law and its steepening on scales smaller than $40$\,$h^{-1}$\,Mpc. Using different releases of the same SDSS NBCKDE photometric catalogue \citet{myers_2006,myers_2007_2} detected only a slight excess in quasar clustering on scales $<35$\,$h^{-1}$\,kpc at $\bar{z}=1.4$. To check the possible deviation of the 2pCF of quasars from a simple power law noted by other 
authors we split the range onto scales $<1'$ and $>1'$ ($\theta=1'$ corresponds to the comoving projected distance of $\sim1.3$\,Mpc at the mean redshift of our \textit{full} sample), but found similar (within $1\sigma$ errors) values of the slope, that agrees with previous results by \citet{myers_2006} who claimed for no evidence for a break at $\sim1'$, based on Poisson statistics. This excess was also studied at $z\sim3-4$ by \citet{Shen_II_2010}, who found that the small-scale amplitude is comparable or lower than power-law extrapolations from the large-scale 2pCF of the SDSS quasars. Obviously, this problem needs more accurate study with larger samples of quasars with spectroscopic redshifts, because modern redshift surveys of quasars lack of close pairs due to the `fibre collision' effect, while the photometric surveys cannot give reliable results on small-scale clustering due to large errors of the photometric redshifts, which are comparable with the redshift difference in close pairs. 


The break in the angular 2pCF, clearly visible by eye at $\sim40'$ corresponds to the comoving projected scale of $\sim50$\,Mpc, and was also previously notices, e.\,g. by \citet{myers_2006}. This break should be related to the fact that the projected 2pCF of quasars at scales $40-50$\,Mpc tends to zero (see e.\,g. \citealt{ivashchenko_2010}), and thus is usually studied up to these scales (e.\,g. \citealt{porciani_2004,daAngela_2005}).

\vspace*{-3ex}
\section{Spatial clustering}\label{sec:4}
\subsection{General notations and technique}\label{sec:4-1}

\indent\indent Similar to the angular 2pCF the spatial 2pCF $\xi(r)$ defines a probability $P_{2}$ to find a neighbour for any quasar at a distance $r$ within the small volume $\delta V$ as:
\begin{equation}
 \delta P = n_{2}\left[1+\xi (r)\right]\delta V,
\end{equation}
where $n_{2}$ is the number volume density of objects. Within some limited distance ranges (approximately $1-50$\,Mpc for quasars) the spatial 2pCF is fitted well with the power law:
\begin{equation}\label{eq:spt_dkf}
\xi(r)=\left(\frac{r}{r_0}\right)^{-\gamma},
\end{equation}
where $r_{0}$ is the correlation length, $\gamma$ is the slope of the correlation function.

\begin{table*}
 \centering
 \begin{minipage}{170mm}
  \caption{The parameters of the  deprojected real-space 2pCF for three samples of quasar candidates comparing with the results of other authors from spectroscopic surveys. Here DR3Q stands for SDSS Quasar Catalog III \citep{schneider+2005},  DR5Q stands for SDSS Quasar Catalog IV \citep{schneider+2007}, $\sigma$ and $r$ are projected and three-dimensional real-space distance. Note, that instead of the mean redshift \citealt{white+2012} indicated the effective one.}\label{tab:r-space}
\centering
  \begin{tabular}{|c|c|c|c|c|c|c|}
  \hline
dist. range, $h^{-1}$Mpc & $z$ range & $z$ & $r_{0}$, h$^{-1}$ Mpc & $\gamma$ &  sample & authors\\
  \hline
			&  & 1.51 (mean) & $7.81^{+1.18}_{-1.16}$ & $1.94\pm0.06$ & \textit{full}  &  \\
$\sigma\approx1-35$   & $0.8-2.2$ & 1.51 (mean) & $6.82\pm1.07$ & $2.20^{+0.11}_{-0.10}$ & \textit{high-reddening}  & this work \\
			& (phot.) & 1.50 (mean) & $9.02\pm1.73$ & $1.96^{+0.08}_{-0.09}$ & \textit{low-reddening}  &  \\
  \hline
$\sigma=3-80$ & $0.8-1.4$ & 1.13 (med.) & $8.16\pm0.79$ & $2$ (fixed) &  SDSS DR5Q & \citealt{Shen_2009} \\
$\sigma=1-130$ & $0.3-2.2$ & 1.27 (med.) & $5.45^{+0.35}_{-0.45}$ & $1.90^{+0.4}_{-0.03}$ &  SDSS DR5Q & \citealt{ross_2009} \\
$r=1-10$  & $0.3-2.2$ & 1.4 (med.) & $6.0^{+0.5}_{-0.7}$ & $1.45\pm0.27$  & 2QZ & \citealt{daAngela_2005}\\
$r=10-40$ & $0.3-2.2$ & 1.4 (med.) & $7.25$ & $2.30^{+0.12}_{-0.03}$  & 2QZ & \citealt{daAngela_2005} \\
$r=1-20$ & $0.8-2.1$ & 1.47 (eff.) & $4.8^{+0.9}_{-1.5}$ & $1.5\pm0.2$ &  2QZ & \citealt{porciani_2004} \\
$\sigma=1.2-30$ & $0.8-2.1$ & $\approx1.45$ & $6.47\pm1.55$ & $1.58\pm.20$  & SDSS DR5 & \citealt{shen_2007} \\
$\sigma=1-35$   & $0.8-2.1$ & $\approx1.47$  (mean.) & $6.95\pm0.57$ & $1.90\pm0.11$ & SDSS DR3Q  & \citealt{ivashchenko_2008_dr3}    \\
$\sigma=1-35$   & $0.8-2.2$ & $\approx1.47$  (mean.) & $5.82\pm0.61$ & $1.91\pm0.11$ & SDSS DR5  & \citealt{ivashchenko_2010_dr5}    \\
$\sigma=1-35$   & $0.8-2.2$ & 1.47  (mean.) & $5.85\pm0.33$ & $1.87\pm0.07$ & SDSS DR7   & \citealt{ivashchenko_2010}    \\
$\sigma=3-80$ & $1.4-2.0$ & 1.68 (med.) & $8.51\pm0.76$ & $2$ (fixed) &  SDSS DR5Q & \citealt{Shen_2009} \\
  \hline
\end{tabular}
\end{minipage}
\end{table*}

The angular 2pCF is a projection of the spatial 2pCF on the celestial sphere, thus one could be recovered from another if the redshift distribution of objects and the cosmological model are known. If the angular and spatial 2pCFs are presented as \eqref{eq:ang_dkf} and \eqref{eq:spt_dkf}, respectively, their parameters are related with the Limber's equation \citep{limber_1953}, from which one can derive:
\begin{multline}\label{eq:limb}
\alpha=\gamma-1,\\
A=H_{\gamma}\frac{\int\left(\frac{dN}{dz}\right)^{2}E_{z}(1+z)^{\gamma-(3+\epsilon)}\chi^{1-\gamma}dz}{\left[\int\frac{dN}{dz}dz\right]^2}r_{0}^{\gamma},
\end{multline}
where $H_{\gamma}=\Gamma(0.5)\Gamma(0.5[\gamma-1])/\Gamma(0.5\gamma)$, $\chi$ is the radial comoving distance, $dN/dz$ is the redshift distribution function of the sample, and $E_{z}=H_{z}/c=dz/d\chi$. $H_{z}$ is the Hubble parameter, which for the spatially flat Universe with $\Lambda$CDM cosmological model has a form:
\begin{equation}
H_{z}^{2}=H_{0}^{2}[\Omega_{m}(1+z)^{3}+1-\Omega_{m}].
\end{equation}
In the linear regime of perturbations growth the expression $(1+z)^{\gamma-(3+\epsilon)}$ in equation \eqref{eq:limb} can be substituted by $D_{z}$ (see e.\,g. \citet{myers_2006} for discussion), where
\begin{equation}
D_{z}=\frac{g_{z}}{g_{0}}\frac{1}{(1+z)}.
\end{equation}
According to \citet{carroll+1992} for the flat $\Lambda$CDM-model $g_{z}$ can be approximated as:  
\begin{equation}
g_{z}=\frac{5}{2}\Omega_{mz}\left[\Omega^{4/7}_{mz}-\Omega_{\Lambda z}+\left(1+\frac{\Omega_{mz}}{2}\right)\left(1+\frac{\Omega_{\Lambda z}}{70}\right)\right]^{-1},
\end{equation}
where the density parameters evolve with redshift as
\begin{equation}
\Omega_{mz}=\left(\frac{H_{0}}{H_{z}}\right)^{2}\Omega_{m}(1+z)^3,\quad \Omega_{\Lambda z}=\left(\frac{H_{0}}{H_{z}}\right)^{2}\Omega_{\Lambda}.
\end{equation}

Having the parameters of the spatial 2pCF of quasars we can found their linear bias parameter, adopting the following representation of the quasar clustering evolution (see e.\,g. \citealt{myers_2006}), in which the matter correlation function is extrapolated to high redshift according to the linear growth model:
\begin{equation}
 \frac{\xi(r,z)}{b^{2}_Q(r,z)}=\xi_{m}(r,z)=\left(\frac{r}{r_{0,m}}\right)^{-\gamma}D_{z}^{2+\gamma},
\end{equation}
and assuming the correlation length of the dark matter to be $r_{0,m}(z=0)=5\,h^{-1}$\,Mpc, which was obtained by \citet{jenkins+1998} in $\Lambda$CDM simulations. In this case the quasar linear bias can be calculated from the following simple expression: 
\begin{equation}\label{eq:bias-est}
b_{Q} = \left(\dfrac{r_{0}}{5h^{-1}\,\text{Mpc}}\right)^{\gamma/2}.
\end{equation}

\subsection{Results and discussion}\label{sec:4-2}

\indent\indent Using the values of the angular 2pCF over the angular scales $1'-40'$ for the {\it full}, {\it low-} and {\it high-reddening} samples and their redshift distributions (Fig.~\ref{fig:z-flh}) the values of $r_{0}$, $\gamma$ were derived from the eq.~\eqref{eq:limb}. Table~\ref{tab:r-space} summarises the obtained parameters compared to previously published results for similar redshifts. Similarly to the angular correlation length, the values of $r_{0}$ for the {\it full}, {\it low-} and {\it high-reddening} samples agree with each other within 1$\sigma$. It is clearly seen from Fig.~\ref{fig:ls-estim-levels} and \ref{fig:acf-red-levels} that the slope and the angular correlation length (and the real-space correlation length on its turn) are strongly correlated, hence some authors (e.\,g. \citealt{Shen_2009,white+2012}) fixed the slope value to be $\gamma=2$, based on the previous results, when fitted the quasar 2pCF.

\begin{table*}
 \centering
 \begin{minipage}{170mm}
  \caption{The values of the quasar bias parameter $b_{Q}$ and the results of other authors for comparison. Here DR5Q stands for SDSS Quasar Catalog IV \citep{schneider+2007}. The numbers in brackets correspond to $\xi(s)/\xi(r)$ method -- [1], and methods proposed by \protect\citet{ivashchenko_2010} -- [2], \protect\citet{croom_2005} -- [3], and \protect\citet{porciani_2004} -- [4], \protect\citet{myers_2006} -- [5]; see explanations in the text.}\label{tab:bias}
\centering
  \begin{tabular}{|c|c|c|c|c|c|c|c|}
  \hline
$z$ range & $z$ & $z$ type & $b_{Q}$ & method & sample & authors\\
  \hline
 & 1.51 & mean phot. & $1.54\pm0.22$ & [5] & \textit{full}  &  \\
$0.8-2.2$ & 1.51 & mean phot. & $1.41\pm0.24$ & [5] & \textit{high-reddening}  & this work \\
 & 1.50 & mean phot. & $1.78\pm0.33$ & [5] & \textit{low-reddening}  &  \\
  \hline
$0.8-1.4$ & 1.13 & median spec. & $2.31\pm0.22$ & [3] & SDSS DR5Q & \citealt{Shen_2009} \\
$0.3-2.2$ & 1.27 & median spec. & $2.06\pm0.03$ & [3] & SDSS DR5Q & \citealt{ross_2009} \\
$0.3-2.2$ & 1.35 & mean spec. & $2.02\pm0.07$  & [3] & 2QZ & \citealt{croom_2005}\\
$0.3-2.2$ & 1.40 & median spec. & $1.50\pm0.20$  & [3] & 2QZ+2SLAQ & \citealt{daAngela_2008} \\
$0.3-2.2$ & 1.40 & median spec. & $2.84^{+1.49}_{-0.57}$ & [1] & 2QZ & \citealt{daAngela_2005} \\
$\approx0.1-2.8$ & $\approx1.4$ & mean phot. & $2.51\pm0.46$ & [5] & SDSS NBCKDE DR1 & \citealt{myers_2006} \\
$0.4-2.3$ & 1.40 & mean phot. & $2.41\pm0.08$ & [3] & SDSS NBCKDE DR4 & \citealt{myers_2007} \\
$0.8-2.1$ & 1.47 & effective phot. & $2.42^{+0.20}_{-0.21}$ & [4] & 2QZ   & \citealt{porciani_2004} \\
$0.8-2.2$ & 1.47 & mean spec. & $1.44\pm0.22$ & [2]  & SDSS DR7 & \citealt{ivashchenko_2010} \\
$1.4-2.0$ & 1.68 & median spec. & $2.96\pm0.26$ & [3] & SDSS DR5Q & \citealt{Shen_2009} \\
\hline
\end{tabular}
\end{minipage}
\end{table*}

It is worth to note here, that additional errors to the deprojected values of $r_{0}$ can be introduced by the uncertainties of photometric redshifts, which are significantly larger than the uncertainties of the spectroscopic ones. Hence the real $dN/dz$ distributions which should be used in the eq.\,\eqref{eq:limb} are broader than the distributions we used. To take this fact into account \citet{brunner+2000} proposed to widen the photometric redshift distribution with the help of two one-tailed Gaussians centred on the endpoints of the redshift interval and with the dispersion equivalent to the dispersion between photometric and spectroscopic redshifts. To estimate the possible error we used the dispersion value of $\sigma=0.3$, which is the estimation of photometric redshift precision of 86\% of objects from the SDSS NBCKDE catalogue calculated by its authors \citep{richards_2004} from comparison with spectroscopic redshifts. After widening the redshift distributions we found the values of $r_{0}$ to be 
systematically larger than those without widening. But as far as the discrepancy is smaller than our 1$\sigma$ errors of $r_{0}$, we present the uncorrected values.

Comparing the results, one can see that our values of $\gamma$, $r_{0}$ for the {\it full} sample are consistent within 1$\sigma$ with results of \citet{ivashchenko_2008_dr3} for the same distance range, with the mean values of two distance ranges from \citet{daAngela_2005}, which in total are similar to our range, as well as  with the results of \citet{Shen_2009} for the both nearest redshift bins ($0.8-1.4$ and $1.4-2.0$), although \citet{Shen_2009} used slightly broader distance range and excluded distances smaller than 3$h^{-1}$\,Mpc. Our values also agree within 2$\sigma$ with the results of \citet{ross_2009}, \citet{ivashchenko_2008_dr3} and  \citet{ivashchenko_2010}, though \citet{ross_2009} used much broader distance range. The exception is the result by \citet{porciani_2004}, who obtained significantly smaller value of $\gamma$, but they used more narrow distance range, and nevertheless these results are still consistent with our values within 2$\sigma$. Note, that \citet{myers_2006} obtained much 
larger value of $r_{0}\approx12$ for the First Release of SDSS NBCKDE catalogue, which can be likely explained by much wider redshift range of the sample, over which the averaging is made. 

The derived values of the bias parameter are summarised in Table~\ref{tab:bias} compared to the previous results by other authors for samples with similar (photometric or spectroscopic) redshift. As far as the technique of determination of the bias parameter is more ambiguous, the values summarised here are obtained with different approaches (we indicate the different methods with numbers in brackets) with spectroscopic samples of quasars, except \citet{myers_2007}, who used photometric catalogue. 

Some of the authors derived $b$ via the parameter $\beta=f(\Omega_{M},z)/b$, characterising the gravitational infall of galaxies onto the matter overdensities. This approach is related to redshift-space distortions and usually suffers from degeneracy between the parameters characterising these distortions ($\beta$ and pairwise velocity dispersion $\left\langle{w^{2}}\right\rangle^{1/2}$) and geometric flattening (usually $\Omega_{M}$), thus in its turn this approach has several variations. The simplest technique, denoted as [1], involves the ratio of the real- and redshift-space 2pCF, $\xi(s)/\xi(r)$, was used by \citet{daAngela_2005}. Similar technique (denoted as [2]) was used by \citet{ivashchenko_2010}, but unlike \citet{daAngela_2005} the authors considered the fitted power-law for both 2pCFs with the same slope (that is valid for monopole part of the 2pCF). Another approach proposed by \citet{croom_2005} (and denoted as [3]) suggests the solution of the quadratic equation in $b_{Q}$ with the shape of $\
xi(r)$ governed by the underlying dark matter distribution described by the analytic equation from \citet{hamilton_1991,hamilton_1995}. \citet{myers_2007}, \citet{daAngela_2008} and \citet{ross_2009} applied the previous approach with the theoretical dark matter power spectrum from \citet{smith_2003}. One more method which was used by \citet{porciani_2004} (denoted as [4]) includes using the relation between projected 2pCFs of the quasars and that of the matter. The simpler technique, described in Sec.\,\ref{sec:4-1} and proposed by \citet{myers_2006}, was used in the present paper (denoted with [5]).
                                                                                                                                                                                                                                                
Our value of $b$ for the {\it full} sample agree well with the values obtained by \citet{daAngela_2008} and \citet{ivashchenko_2010}. The values obtained by other authors are significantly larger. Nevertheless, our value is still consistent within 2$\sigma$ with \citealt{croom_2005}, \citealt{daAngela_2005} and within 3$\sigma$ with \citealt{porciani_2004}, \citealt{myers_2006,myers_2007}, \citealt{Shen_2009}, \citealt{ross_2009}.  

When comparing the values of $r_{0}$, $\gamma$ and $b(z)$ from Tables\,\ref{tab:r-space}, \ref{tab:bias}, one has to remember, that besides the differences in techniques and samples, which in general should not influence the result and should serve as an additional verification of the results obtained by different authors, there are some other aspects which can cause the discrepancy. The first one, which is more subjective, is the fact that different authors indicate different general characteristics of the samples they used. Some authors indicate the mean redshift, while other use the median or effective redshift as the characteristics of the sample. It is obvious, that these three values are not equivalent for samples with typical non-uniform redshift distribution of quasars. For example, in Table~\ref{tab:zcomp} we present the values of the mean, median and effective redshift for four of our samples: {\it full}, {\it nearby}, {\it medium-z} and {\it distant}. Here we derived the effective redshift 
integrating over the redshift distribution as $z_{eff}=\left[\int\left(dN/dz\right)^{2}zdz\right]\cdot\left[\int\left(dN/dz\right)^{2}dz\right]^{-1}$. Note, that different authors (see e.\,g. \citealt{porciani_2004,white+2012}) use different definitions of $z_{eff}$.  One can see, that for the samples, the redshift distributions of which are far from symmetric, like {\it nearby} and {\it distant}, there is a discrepancy between these three values. Taking this aspect into account, the redshift interval seems to be more suitable sample characteristics in case of comparison. That is why in both tables we indicate the type of the redshift value used by authors as a general characteristics, and the redshift interval.

\begin{table}
 \centering
 \begin{minipage}{80mm}
\caption{Mean, median and effective $z$ for some samples.}\label{tab:zcomp}
\centering
  \begin{tabular}{|c|c|c|c|}
  \hline
sample & $z_{mean}$ & $z_{med}$ & $z_{eff}$ \\
  \hline
{\it full} & 1.51 & 1.48 & 1.50 \\
{\it nearby} & 0.89 & 1.00 & 1.08 \\
{\it medium-z} & 1.68 & 1.69 & 1.73 \\
{\it distant} & 2.63 & 2.49 & 2.55 \\
\hline
\end{tabular}
\end{minipage}
\end{table}

The second, more objective, aspect is the distance range within which the 2pCF is fitted. It is known, that the 2pCF of quasars is described well with a single power law only within a limited distance range. The upper limit is usually restricted by a sample volume and statistics, while the lower limit, besides these facts, is also related to the scales where the transition from linear to non-linear regime occurs. It is known that the pure linear scales are approximately $>10\,h^{-1}$\,Mpc, and when deriving the bias parameter via $\beta$ these scales are usually used (see e.\,g. \citealt{daAngela_2005}). In case of other techniques the smaller scales up to $\sim1\,h^{-1}$\,Mpc are used.

\section{Luminosity and z-dependence}\label{sec:5}

\subsection{Results}\label{sec:5-1}

\begin{figure*}
\centering
\begin{minipage}{.95\textwidth}
\centering
\epsfig{figure=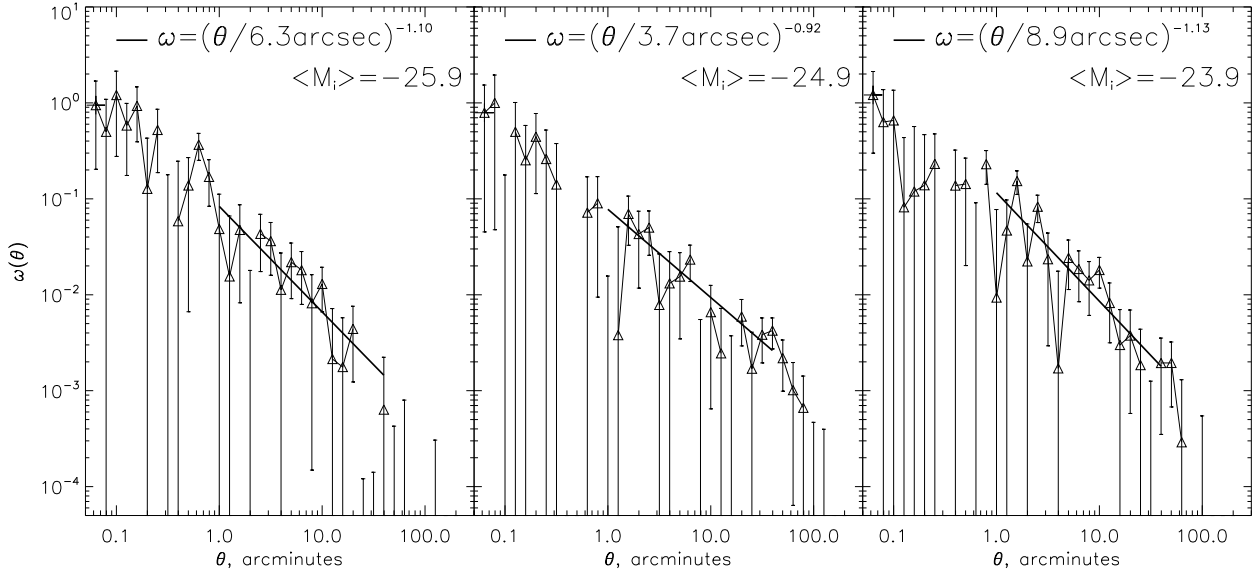, width=.99\linewidth}
\caption{The angular 2pCF for (left to right) \textit{bright}, \textit{medium-M} and \textit{faint} quasars. The solid line is the best fit within $1'-40'$ angular range.}\label{fig:lum-fit}
\end{minipage}
\end{figure*}

\begin{figure*}
\centering
\begin{minipage}{.95\linewidth}
\centering
\epsfig{figure=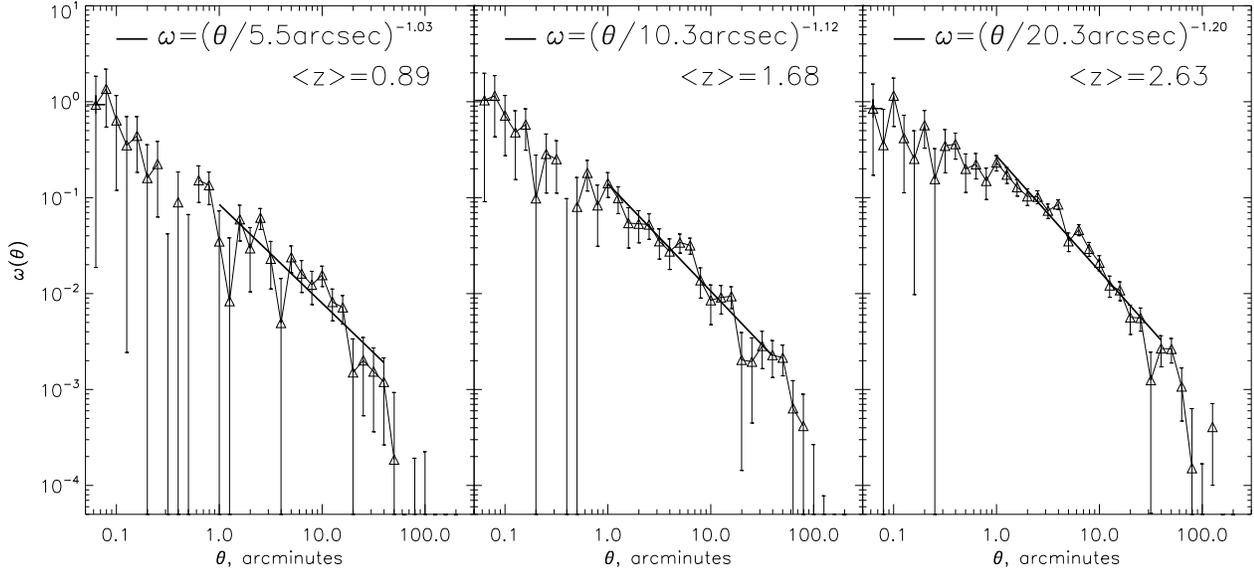, width=.99\linewidth}
\caption{The angular 2pCF for (left to right) \textit{nearby}, \textit{medium-z} and \textit{distant} quasars. The solid line is the best fit within $1'-40'$ angular range.}\label{fig:evo-fit}
\end{minipage}
\end{figure*}

\begin{figure}
\centering
\begin{minipage}{.45\textwidth}
\centering
\epsfig{figure=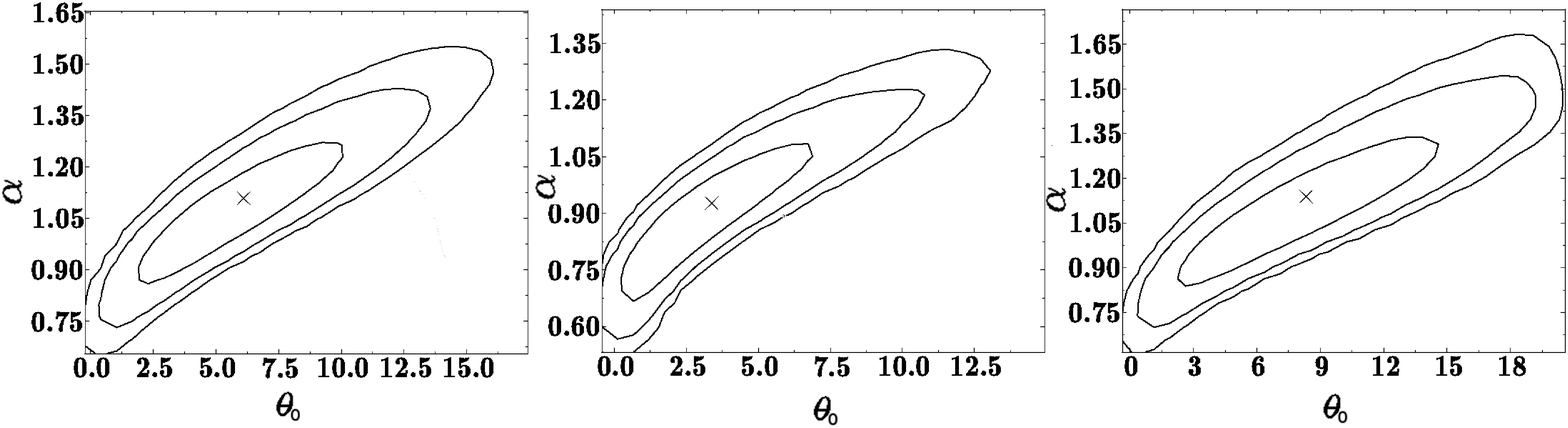, width=.99\textwidth}
\caption{1,\,2,\,3$\sigma$ confidence levels for $\alpha$ and $\theta_{0}$ parameters of the angular 2pCF for the samples of (left to right) \textit{bright-}, \textit{medium-M} and \textit{faint} quasars samples within $1'-40'$ angular range.}\label{fig:lum-params}
\end{minipage}
\end{figure}

\begin{figure}
\centering
\begin{minipage}{.45\textwidth}
\centering
\epsfig{figure=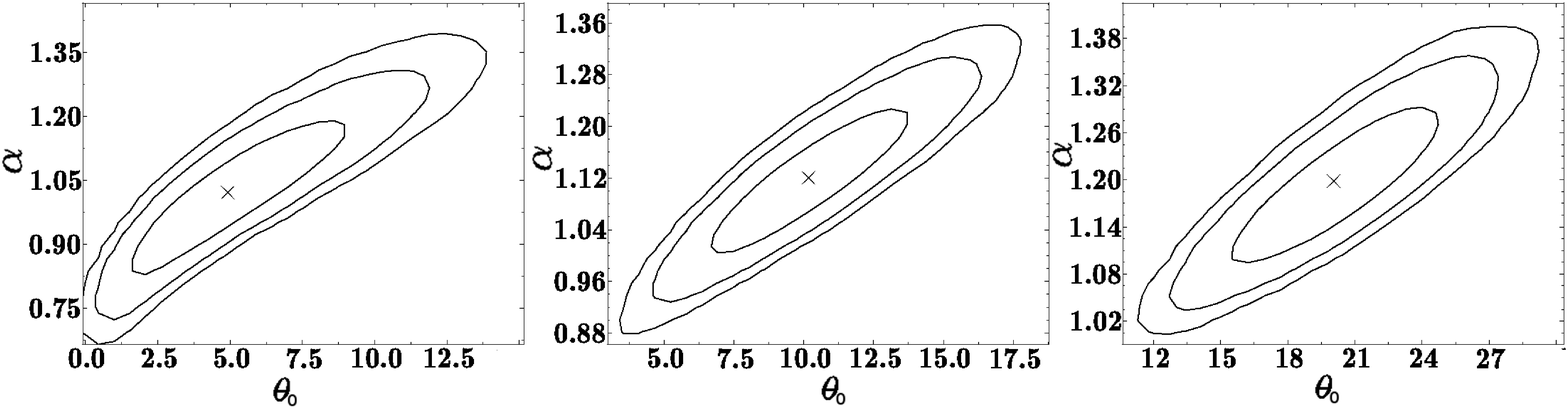, width=.99\textwidth}
\caption{1,\,2,\,3$\sigma$ confidence levels for $\alpha$ and $\theta_{0}$ parameters of the angular 2pCF for the samples of (left to right) \textit{nearby}, \textit{medium-z} and \textit{distant} quasars samples within $1'-40'$ angular range.}\label{fig:evo-params}
\end{minipage}
\end{figure}

\indent\indent In Fig.~\ref{fig:lum-fit} and \ref{fig:lum-params} the obtained 2pCF for the samples with different luminosity along with the best fits within $1'-40'$ range and the likelihood function levels corresponding to 1, 2, and 3-$\sigma$ marginalised errors of $\alpha$ and $\theta_{0}$ are shown. The same results for the samples with different redshift are presented in Fig.~\ref{fig:evo-fit}, \ref{fig:evo-params}. The best fit values of the angular 2pCF parameters along with derived parameters of the spatial 2pCF and $b$-parameters are summarised in Table~\ref{tab:lum-params}. 

In Fig.~\ref{fig:r0} and \ref{fig:bias} the redshift dependence of the correlation length $r_{0}$ and the quasar bias parameter according to the previous results is shown along with our values for the {\it nearby}, {\it medium-z} and {\it distant} samples. 

\subsection{Comparison with optical quasar clustering}\label{sec:5-2}

\indent\indent The luminosity dependence of the quasar clustering was previously studied mainly for the 2QZ and the SDSS surveys, the large size of which allows to divide a sample onto redshift and luminosity bins simultaneously. \citet{Croom+2002} studied the quasar clustering as a function of luminosity over the redshift range $0.3<z<2.9$ using a sample from the preliminary data release of the 2QZ catalogue, and analysing the integrated 2pCF over scales $<20\,h^{-1}$\,Mpc found only a marginal evidence for quasars with brighter apparent magnitudes having a strong clustering amplitude. Later, using the final 2QZ catalogue in combination with the 6QZ catalogue \citet{croom_2005} found no significant evidence for any dependence of the quasar clustering amplitude on apparent magnitude over $0.3<z<2.2$ range. \citet{porciani_2006} also used a sample of quasars from the final edition of the 2QZ catalogue with $0.8<z<2.1$ and applying 5 different methods also found a marginal evidence for luminosity dependent 
clustering for $z>1.3$, but noted that the sample is too small to provide a robust detection.

The most detailed analysis of possible luminosity dependence of the quasar clustering from the SDSS spectroscopic sample was carried out by \citet{Shen_2009}, who used a sample of quasars from the SDSS Data Release Five quasar catalogue within $0.4<z<2.5$, and found that at $z<2.5$ quasar clustering depends weakly on luminosity, namely the most luminous and most massive quasars are more strongly clustered (at the $\sim2\sigma$ level) than other quasars. The lower redshift ($0.2<z<0.6$) quasars clustering in the same release of the SDSS was studied by \citet{padmanabhan+2009}, who found no evidence for its luminosity evolution. The higher redshift range ($2.4<z<2.8$) was investigated recently by \citet{white+2012} with the BOSS survey data. They also have not found any evidence for luminosity dependence of the clustering amplitude, but noted, that the sensitivity of the given sample for this dependence is weak due to the limited dynamical range of the sample in luminosity.

\begin{table*}
\centering
\begin{minipage}{.95\textwidth}
\caption{Parameters of the angular and spatial 2pCF, and the bias parameter for the samples of quasars with different luminosity and redshift within $1'\leq\theta\leq40'$ angular range.}\label{tab:lum-params}
\centering
\vspace*{1ex}
\begin{tabular}{|c|c|c|c|c|c|c|}
 \hline
sample & $\alpha$ & $\theta_{0}$, arcsec & $\chi^{2}/$d.o.f. & $\gamma$ & $r_{0}$, $h^{-1}$\,Mpc & $b$ \\
\hline
{\it bright} & $1.10^{+0.12}_{-0.14}$ & $6.3\pm2.8$ & 0.8 & $1.20\pm^{+0.12}_{-0.14}$ & $7.20\pm1.68$ & $1.47\pm0.36$ \\
{\it medium-M} & $0.92^{+0.13}_{-0.14}$ & $3.7\pm2.4$ & 1.2 & $1.92^{+0.13}_{-0.14}$ & $7.63\pm2.3$ & $1.50\pm0.44$ \\
{\it faint} & $1.13\pm0.16$ & $8.9\pm4.2$ & 0.8 & $2.13\pm0.16$ & $5.05^{+1.26}_{-1.25}$ & $1.01\pm0.27$\\
\hline
\textit{nearby} & $1.03^{+0.12}_{-0.11}$ & $5.5\pm2.5$ & 1.6 & $2.03^{+0.12}_{-0.11}$ & $4.56\pm1.03$ & $0.91\pm0.21$ \\
\textit{medium-z} & $1.12^{+0.08}_{-0.07}$ & $10.3^{+1.5}_{-2.4}$ & 2.4 & $2.12^{+0.08}_{-0.07}$ & $7.16^{+0.56}_{-0.87}$ & $1.46^{+0.12}_{-0.19}$ \\
\textit{distant} & $1.20^{+0.06}_{-0.07}$ & $20.3\pm3.0$ & 1.1 & $2.20^{+0.06}_{-0.07}$ & $13.62^{+1.08}_{-1.07}$ & $3.01\pm0.26$ \\
\hline
\end{tabular}
\end{minipage}
\end{table*}


\begin{figure*}
\centering
\epsfig{figure=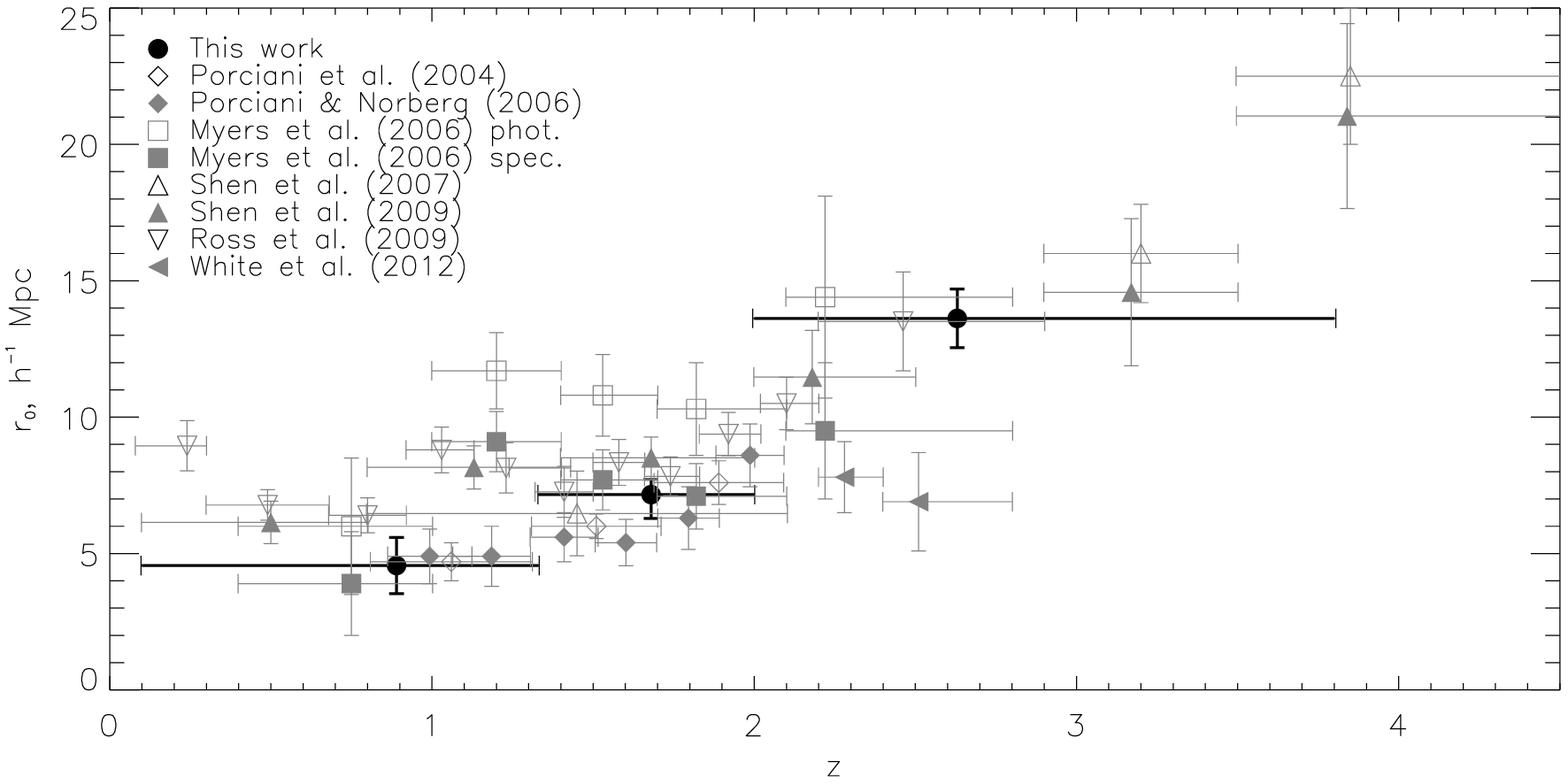, width=.95\linewidth}
\caption{Redshift-dependence of the real-space correlation length in comparison with results of previous studies. The points show the results obtained from 2QZ catalogue by \protect\citet{porciani_2004} and \protect\citet{porciani_2006}, photometrically selected quasar candidates and spectroscopically confirmed quasars from SDSS NBCKDE (DR1) by \protect\citet{myers_2006}, SDSS DR5 by \protect\citet{shen_2007}, SDSS Quasar Catalog IV by \protect\citet{Shen_2009}, \protect\citet{ross_2009}, and BOSS (SDSS DR9) by \protect\citet{white+2012}. Error-bars in $z$ show the redshift range, over which the 2pCF was averaged.}\label{fig:r0}
\end{figure*}

\begin{figure*}
\centering
\epsfig{figure=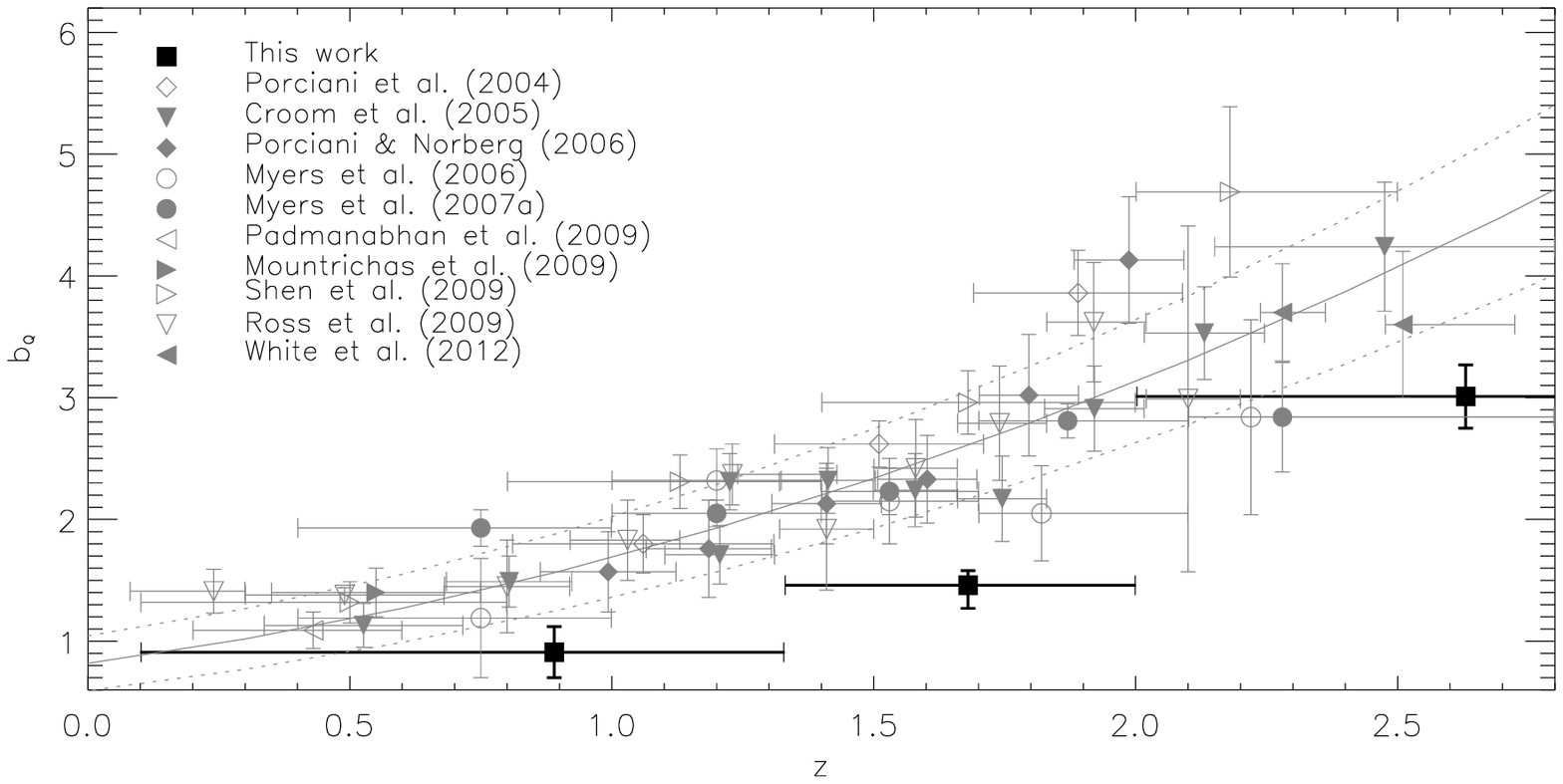, width=.95\linewidth}
\caption{Redshift-dependence of the bias parameter in comparison with results of previous studies. The solid and dashed lines denote a semi-empiric relation $b_{Q}(z)=0.53+0.289(1+z)^2$ and its 1$\sigma$ error range derived by \protect\citet{croom_2005} from 2QZ catalogue. The points show the results obtained from 2QZ catalogue by \protect\citet{croom_2005}, \protect\citet{porciani_2004} and \protect\citet{porciani_2006}, photometrically selected quasar candidates from SDSS NBCKDE (DR1 and DR4) by \protect\citet{myers_2006,myers_2007}, SDSS Quasar Catalog IV by \protect\citet{padmanabhan+2009}, \protect\citet{Shen_2009} and \protect\citet{ross_2009}, 2SLAQ + 2QZ + SDSS DR5 + AAOmega surveys by \protect\citet{Mountrichas_2009}, and BOSS (SDSS DR9) by \protect\citet{white+2012}. Error-bars in $z$ show the redshift range, over which the result was averaged.}\label{fig:bias}
\end{figure*}

A flux limitation of any redshift survey results in the fact, that more luminous quasars lie at higher redshifts, thus it is always hard to construct samples with broad luminosity range within a narrow redshift bin. \citet{daAngela_2008} used a combination of the 2QZ and fainter 2SLAQ (2dF-SDSS LRG and QSO catalogue, \citealt{richards_2005}; LRG is a luminous red galaxy) quasar surveys, that allowed them to widen the magnitude window and study luminosity dependence within $0.3<z<2.9$ range, free of any evolutionary effects. Their results are consistent with luminosity-independent clustering.

As one can see from Table\,\ref{tab:lum-params}, our values of $r_{0}$, $\gamma$ and $b$ for {\it faint} sample differ from those for {\it bright} and {\it medium-M} samples, but nevertheless, the values for all three samples with different luminosity are consistent within 1$\sigma$. The discrepancy can be attributed to the difference in redshift distribution for these samples, discussed in Sec.\,\ref{sec:2-4}. Hence, we can not argue for any luminosity dependence of the quasar clustering. This result is consistent with previous studies of spectroscopically confirmed quasars, as far as with study of similar samples of photometrically classified quasars by \citet{myers_2006,myers_2007}. For the Early Release of the SDSS NBCKDE catalogue \citet{myers_2006} have not found any evidence in luminosity dependence, while using larger sample \citet{myers_2007} argued for $1.5\sigma$ deviation from luminosity independent quasar clustering at $\bar{z}\sim1.9$. 

It is also seen from Table\,\ref{tab:lum-params} that the correlation length $r_{0}$ increases substantially with the redshift, while the slope remains almost constant. The increase in $r_{0}$ with redshift is clearly seen in Fig.\,\ref{fig:r0}, where we plotted our results along with the results of previous studies, which show the same trend. The constant shape of the 2pCF was previously noted by several authors. E.\,g. \citet{croom_2005} found the shape of the redshift-space 2pCF to be constant, testing changes both in the ratios $\bar{\xi}_{s}(20)/\bar{\xi}_{s}(30)$ and $\bar{\xi}_{s}(50)/\bar{\xi}_{s}(30)$, and the slope $\gamma$ within $0.3<z<2.9$. Similar result was obtained by \citet{porciani_2004} and \citet{porciani_2006} for the real-space 2pCF for the same 2QZ catalogue within $0.8<z<2.1$, thus the values of $r_{0}$ from \citealt{porciani_2004} and \citealt{porciani_2006} shown in Fig.\,\ref{fig:r0} are those obtained for a fixed $\gamma=1.8$. On the other hand, if the subsamples are small for a 
flexible power-law-index fit, the fixed values of $\gamma$, close to the observed values, are also used. E.\,g. the values from \citealt{myers_2006}, presented in Fig.\,\ref{fig:r0}, for photometric data from the SDSS NBCKDE catalogue are obtained with $dN/dz$ widened by Gaussians for a fixed $\gamma=1.98$, the same $\gamma$ was used for projected 2pCF of spectroscopically confirmed quasars from this catalogue within $0.75-35\,h^{-1}$\,Mpc. The values from \citealt{Shen_2009} and \citealt{ross_2009} are obtained for fixed $\gamma=2$ for all and low-redshift objects in SDSS DR5 quasar catalogue \citep{schneider+2007}, respectively; the same $\gamma=2$ was used by \citet{white+2012} for the sample of high-z quasars from the BOSS survey. Only the values of $r_{0}$ from \citealt{shen_2007} are taken from the variable power-law index fit (see Table\,\ref{tab:r-space}) for high-redshift quasars from SDSS DR5, but the authors note that they are consistent with those got with fixed $\gamma=2$.    

\begin{table*}
 \centering
\begin{minipage}{.95\textwidth}
\centering
\caption{Parameters of the 2pCF and bias parameter of X-ray sources from \protect\citealt{Gilli+2005} (G05), \protect\citealt{Basilakos+2005} (B05), \protect\citealt{Yang+2006} (Y06), \protect\citealt{Miyaji+2007} (M07), \protect\citealt{ebrero_2009} (E09), \protect\citealt{coil+2009} (C09), \protect\citealt{Gilli+2009} (G09), \protect\citealt{elyiv_2011} (E11), \protect\citealt{Allevato+2011} (A11).} \label{tab:xray}
 \vspace*{2ex} 
\begin{tabular}{c|c|c|c|c|c|c|c|}
 \hline
  survey & band, keV & objects & $\bar{z}$ & slope $\gamma$ & $r_{0},\,h^{-1}$Mpc & $b$ & authors \\
 \hline
 CDF-N & 0.5-10 & AGN I & 1.03 (med.) & $1.46\pm0.33$ & $9.1\pm3.3$ & -- & \\
 CDF-N & 0.5-10 & AGN II & 0.73 (med.) & $1.40\pm0.21$ & $10.5\pm2.2$ & -- & G05 \\
 CDF-S & 0.5-10 & AGN I & 1.02 (med.) & $1.89\pm0.23$ & $6.5\pm0.8$ & -- & \\
 CDF-S & 0.5-10 & AGN II & 0.87 (med.) & $1.52\pm0.27$ & $5.1\pm1.3$ & -- & \\
 \hline
 XMM-Newton/2dF & 0.5-2 & -- & 1.2 (med.) & -- & $\sim16$ ($\gamma=1.8$) & $1.9-2.7$ & B05 \\
 \hline
 CDF-N & 0.5-10 & all AGN & 0.84 (med.) & $1.38^{+0.12}_{-0.14}$ & $4.1^{+0.7}_{-1.1}$ & -- & \\ 
 CLASXS & 2-8 & all AGN  & 1.2 (med.) & $2.1\pm0.5$ & $5.8^{+0.9}_{-1.6}$ & -- & Y06 \\
 CDF-N+CLASXS &  & all AGN  & 0.94 (med.) & 1.47 & 4.3 & $2.04\pm1.02$ &  \\
 \hline
 XMM-COSMOS & 0.5-2 & -- & 1.07 (eff.) & -- & $11.8\pm1.1$ ($\gamma=1.8$) & $3.7\pm0.3$ &  \\
 XMM-COSMOS & 2-4.5 & -- & 0.87 (eff.) & -- & $6.9^{+2.2}_{-3.1}$ ($\gamma=1.8$) & $2.5^{+0.7}_{1.0}$ &  M07\\
 XMM-COSMOS & 4.5-10 & -- & 0.60 (eff.) & -- & $12.7^{+2.3}_{-2.7}$ ($\gamma=1.8$) & $3.8^{+0.6}_{-0.8}$ &  \\
 \hline
 XMM & 0.5-2 & -- & 0.96 (med.) &  $2.12\pm0.04$ &  $12.25\pm0.12$ & $4.82\pm0.18$ & \\ 
 XMM & 2-10 & -- & 0.94 (med.) & $2.33^{+0.10}_{-0.11}$ & $ 9.9\pm2.4$ & $4.53\pm1.06$ & E09\\ 
 XMM & 0.5-4.5 & -- & 0.77 (med.) & $2.47^{+0.43}_{-0.57}$ & $7.0\pm5.5$ & $3.16\pm2.97$ & \\  
 \hline
 AEGIS & 0.5-7 & non-QSO & 0.90 (med.) & $1.66\pm0.22$ & $5.95\pm0.90$ & -- & C09 \\ 
 \hline
 XMM-COSMOS & 0.5-10 & all AGN & 0.98 (med.)  & $1.88^{+0.06}_{-0.07}$ & $8.65^{+0.41}_{-0.48}$ & $2.0\pm0.2$ & \\
 XMM-COSMOS & 0.5-10 & BLAGN & 1.45 (med.) & $1.98^{+0.17}_{-1.18}$ & $7.66^{+0.81}_{-1.84}$ & -- & G09 \\
 XMM-COSMOS & 0.5-10 & non-BLAGN & 0.70 (med.) & $1.60^{+0.13}_{-0.14}$ & $7.03^{+0.87}_{-1.18}$  & -- & \\
 \hline
 XMM-LSS & 0.5-2 & -- & 1.1 (med.) & $1.93\pm0.03$ & $6.2\pm0.7$ ($\gamma=1.8$) & $2.2\pm0.2$ & \\ 
 XMM-LSS & 2-10 & -- & 1.0 (med.) & $1.98\pm0.04$  & $10.1\pm0.9$ ($\gamma=1.8$) & $3.3\pm0.3$ & E11\\ 
 \hline 
 XMM-COSMOS & 0.5-2 & all AGN & 1.22 (med.) & $1.81^{+0.04}_{-0.04}$ & $7.12^{+0.28}_{-0.18}$ & $2.98\pm0.13$ &  \\
 XMM-COSMOS & 0.5-2 & BLAGN & 1.55 (med.) & $1.88^{+0.04}_{-0.06}$ & $7.08^{+0.30}_{-0.28}$ & $3.43\pm0.17$ &  \\
 XMM-COSMOS & 0.5-2 & non-BLAGN & 0.74 (med.) & $1.69^{+0.05}_{-0.05}$ & $7.12^{+0.22}_{-0.20}$ & $2.70\pm0.22$ &  A11\\
 XMM-COSMOS & 0.5-2 & X-unobs. & 1.12 (med.) & -- & -- & $3.01\pm0.21$ &  \\
 XMM-COSMOS & 0.5-2 & X-obs. & 1.30 (med.) & -- & -- & $1.80\pm0.15$ &  \\
 \hline
\end{tabular}
\end{minipage}
\end{table*}
                                                                                                                                                                                                                                                                                                                                                                                      
To compare our results on redshift dependence of the quasar bias we plotted them in Fig.\,\ref{fig:bias} along with the previous results obtained by other authors. As one can see from Fig.\,\ref{fig:r0} and \ref{fig:bias}, our results on $r_{0}$ agree well with the results of previous studies of spectroscopic quasar samples, unlike e.\,g. \citet{myers_2006} who did not place such restrictions on the probability of redshift determination, as we do, that can result in substantial `contamination' of low-redshift samples with high-redshift quasars, and hence in overestimation of $r_{0}$. On the other hand, agreement of our $b_{Q}$ values with the empiric approximation from \citealt{croom_2005}, with which most of other results are consistent, is only within 2$\sigma$. The possible reasons of this discrepancy were discussed in Sec.\,\ref{sec:4-2}.  

\subsection{Comparison with X-ray AGN clustering}\label{sec:5-3}

\indent\indent We also compare our clustering measurements with the results from deep X-ray surveys, which now allow to study the clustering up to $z\approx1-1.5$. We tried to summarise their recent results for the highest redshifts, which are close to our samples, in Table\,\ref{tab:xray}. About 95\% of X-ray point-like sources away from the galactic plane are usually considered to be AGNs. These AGNs are largely missed in optical surveys due to their small optical luminosity. X-ray sources with high hardness ratio of the spectrum are considered to be obscured and hence belong to type II AGNs which do not have broad emission lines in optical range, like Seyfert 2 galaxies, while the soft-spectrum sources are mostly unobscured (type I) AGNs. Generally speaking, this classification is not strict, e.\,g.  \citet{Gilli+2009} noted that the class of non-broad line AGNs (no-BLAGN) may include obscured AGN, weak unobscured AGN, whose optical emission is diluted by the host galaxy light, and normal galaxies. 
Therefore the objects classification in Table\,\ref{tab:xray} is given according to authors. The type I/type II division in \citealt{Gilli+2005} was made using the hardness of X-ray spectrum, while \citet{Gilli+2009} made this division using optical spectroscopic data. \citet{coil+2009} selected non-quasar objects with the help of optical colour-magnitude diagram for the AGN host galaxies. More accurate classification was made by \citet{Yang+2006} and \citet{Allevato+2011}, who used both X-ray spectrum hardness and optical spectroscopy data. E.\,g. \citet{Yang+2006} found that the fractions of type I AGNs in the soft- and hard-spectrum subsamples used by them are 56.4\% and 15.4\%, respectively. In cases when additional classification was not provided by authors, most part of objects are likely AGNs.

Due to small optical luminosity the cross-identification of X-ray sources with the optical counterparts, and hence the redshift measurement, is a complicated task. That is why the same technique of deprojection of the angular 2pCF is often used to study the X-ray sources clustering. This technique allows to use large samples, but its results are dependent on the choice of the clustering model and luminosity function, which is needed to derive the redshift distribution. E.\,g. \citet{Basilakos+2005} studied the angular 2pCF of 432 objects in soft $0.5-2$\,keV band and 462 objects in total $0.5-8$\,keV band from XMM-Newton/2dF survey and deprojected it using the luminosity-dependent density evolution (LDDE) model for the X-ray luminosity function from \citealt{Miyaji+2000}. \citet{Miyaji+2007} conducted the first measurements of the angular 2pCF of X-ray point sources detected in the XMM-Newton observations of the 2\,deg$^{2}$ Cosmic Evolution Survey (COSMOS) field and deprojected it using the LDDE model from \
citealt{Ueda+2003}. Later, \citet{ebrero_2009} measured the angular 2pCF of $\sim30\,000$  X-ray sources observed by XMM-Newton and found that obscured and unobscured objects share similar clustering properties. Therefore they both reside in similar environments, that is in agreement with the unified model of AGN. Recently, \citealt{elyiv_2011} measured the angular 2pCF of $\sim8\,200$ X-ray sources from XMM-Newton Large Scale Structure (XMM-LSS) survey. They argue that the sources with a hard spectrum are more clustered than the soft-spectrum ones, that may evidence for these types of sources to reside in different environments. The values of $r_{0}$ from \citealt{ebrero_2009} and \citealt{elyiv_2011} presented in Table\,\ref{tab:xray} were obtained for LDDE model from \citealt{ebrero_2009_a}. The values of $r_{0}$, $b$ from \citealt{Basilakos+2005,Miyaji+2007,ebrero_2009,elyiv_2011}, presented in Table\,\ref{tab:xray} are obtained within the clustering evolution model, which is constant in comoving 
coordinates, i.\,e. for $\epsilon=\gamma-3$. Note, that the median or effective $z$ indicated for the works mentioned above are also derived from the model.

On the other hand, with the appearance of the deep multi-wavelength survey it become possible to study the 3-dimensional clustering of X-ray sources directly. E.\,g. \citet{Gilli+2005} measured the projected correlation function over scales $\sim0.2-10\,h^{-1}$\,Mpc for a sample of 240 sources with spectroscopic $z$ in the 2 Ms Chandra Deep Field North (CDF-N) and 124 sources in the 1 Ms Chandra Deep Field South (CDF-S). They have detected a significantly different clustering amplitude for 2 fields, but argued, that it is likely due to cosmic variance. \citet{Yang+2006} calculated both redshift-space and projected 2pCF of 233 X-ray point sources from the Chandra Large Area Synoptic X-Ray Survey (CLASXS) and 252 sources the CDF-N, the redshifts of which were measured using follow-up spectroscopy at Keck and Subaru telescopes. The comoving scales used for fitting the projected 2pCF were $<20$\,Mpc and $<30$\,Mpc for CDF-N and CLASXS, respectively. Using the $\xi(s)/\xi(r)$ relation they have also found the 
bias parameter for the combined sample. Note, that the median redshift of the CDF-N sample presented in the Table\,\ref{tab:xray} for \citealt{Yang+2006} was taken from \citealt{Gilli+2005}, who used the same sample, and the values of $r_{0}$ were corrected for $h^{-1}$ with $h=0.71$ used by \citealt{Yang+2006}, as they are presented in Mpc by authors. \citet{Yang+2006} found no significant difference in clustering between hard- and soft-spectrum sources. The spatial clustering of 538 X-ray sources from XMM-COSMOS survey at $z=0.2-3.0$ was studied by \citet{Gilli+2009}. They found no evidence for AGN with broad optical lines to cluster differently from AGN without broad optical lines, and concluded that obscured and unobscured AGN are consistent with inhabiting similar environments. With the technique similar to that used in the present work, they also estimated the bias parameter for the full sample. The redshift cut at $z>0.2$ was applied by \citet{Gilli+2009} for the non-BLAGN sample to guarantee that it 
is mostly populated by AGN in which the absence of broad optical lines is solely due to nuclear obscuration. Recently, \citet{Allevato+2011} studied the redshift evolution of the projected 2pCF of 593 X-ray selected AGNs with spectroscopic redshifts $z<4$, extracted from the 0.5-2\,keV X-ray mosaic of COSMOS, and using different techniques found an evidence of a redshift evolution of the bias factor for the total population of XMM-COSMOS AGNs. After separate study of subsamples of AGNs with and without broad optical lines, and X-ray unobscured and obscured AGNs, they concluded that, for moderate-luminosity X-ray selected AGNs with broad optical lines, the contribution from major mergers is outnumbered by other processes, possibly secular ones such as tidal disruptions or disk instabilities. The values of $b$ from \citealt{Allevato+2011} presented in Table\,\ref{tab:xray} were calculated in the halo model approach. 
 
The X-ray samples are 1-2 orders smaller than the optical ones, thus for increasing of statistics the cross-correlation function with more abundant objects is also used. E.\,g. \citet{coil+2009} measured the cross-correlation function of 113 Chandra-selected X-ray non-quasar AGNs at $z = 0.7-1.4$ with $\sim5\,000$ DEEP2 galaxies in AEGIS field, and found that X-ray AGN have a similar clustering amplitude as red, quiescent and ``green'' transition galaxies at $z\sim1$ and are significantly more clustered than blue, star-forming galaxies. \citet{coil+2009} argue that the X-ray AGN clustering strength is primarily determined by the host galaxy colour, but does not depend on optical brightness, X-ray luminosity, or hardness ratio, at least within the ranges probed in that study. Comparing their results with those of \citet{coil_2007} the authors conclude that high-accretion AGN, observed as optically bright quasars, are less clustered and thus are likely hosted in star-forming galaxies, while X-ray AGN are 
hosted both by massive blue galaxies and red galaxies.

Our values of $r_{0}$, $\gamma$ for the {\it full} sample agree well with the results of \citet{Gilli+2009} and \citet{Allevato+2011} for the samples of BLAGN with similar redshift, that is expected because quasars are a part of the type I AGNs (AGNs with broad emission lines), while the slopes for non-BLAGN samples in both papers are smaller than for BLAGN and for quasars from our sample. The same good agreement (within $\approx1\sigma$) is seen for the {\it nearby} sample ($\bar{z}=0.89$) with the CDF-S AGN type I sample ($\bar{z}=1.02$) from \citealt{Gilli+2005}, but due to large errors of that results and discrepancy between CDF-N and CDF-S sample a comparison with results by \citet{Gilli+2005} is complicated. Anyway the values of $\gamma$ for samples of type II AGNs in \citealt{Gilli+2005} are smaller than those for type I AGNs, although they are consistent within errors. The lower value of $\gamma$ was also obtained for non-quasar objects by \citet{coil+2009}.

For those samples, which contain a mixture of X-ray point sources a comparison should be done mostly with soft band sources ($<2$\,keV), because of obscuration which decreases with increasing of X-ray energy. Thus the sources seen in soft band are probably unobscured and related to type I AGNs. Although a good agreement is seen between values of $r_{0}$, $\gamma$ for the {\it nearby} sample only with soft sample ($z=1.1$) from \citealt{elyiv_2011}. As for other studies of ``mixed'' samples \citep{Yang+2006,Miyaji+2007,ebrero_2009}, our values of 2pCF parameters are closer to those for hard band sources.   

Our values of the bias parameter are systematically lower than those for X-ray samples with similar redshifts, which agree with the results of previous studies (see e.\,g. \citealt{Allevato+2011}). Note, that for lower redshifts optically and X-ray selected AGNs are found to have similar clustering strength (see e.\,g. \citealt{krumpe+2010}) 

\subsection{Discussion}\label{sec:5-4}

\indent\indent According to the standard picture of gravitational growth of structure, the mass distribution evolves with redshift and objects should be more clustered in later epoch. On the other hand, as it is clearly seen from Fig.\,\ref{fig:r0}, the quasars appear to be more clustered in earlier epoch. It means that they are biased tracers of the matter distribution and the bias evolves with time, as it is shown in Fig.\,\ref{fig:bias}. The evolution of the linear bias parameter can be used, e.\,g. for estimation of the typical mass of dark matter halo hosting quasars with the theoretical relations derived by \citet{mo+1996}, \citet{Sheth+2001}. It was shown by many authors (see e.\,g. \citealt{croom_2005}), that the host halos of quasars have the same average mass at all redshift, on the other hand the mass of the dark matter halos should grow with time. It places limits on the lifetime of the quasar activity, which can be estimated within a given model (e.\,g. \citealt{haiman+01,martini+01}). And 
finally, having the mass of the dark matter halo and assuming its density profile, one can determine the mass of the central black hole from the relations, suggested by \citet{Ferrarese_2002}, \citet{Wyithe+2005}. 

It is believed that the main mechanism, that drives the formation of supermassive black holes and triggers quasar activity in ultraviolet and X-ray range, is galaxy mergers. The first simple semi-analytic models of quasar formation (e.\,g. \citealt{kauffmann+2000}, \citealt{wyithe+2003}) assume that quasars radiate at fixed luminosity for some characteristic lifetime, thus there should be a tight relation between the instantaneous quasar luminosity and the mass of the central black hole, which in its turn depends on the dark matter halo mass. These models are completely ruled out with observations, that show the absence of strong luminosity dependence of the quasar clustering. The next generation of models (e.\,g. \citealt{hopkins+2005,hopkins+2005_2}, \citealt{lidz+2006}) assumes that bright and faint quasars are similar sources, observed at different stages of their evolution, thus a broad range of the quasar luminosity corresponds to a narrow range in the masses of quasar host halos, and quasar clustering 
should depend only weakly on luminosity. Predictions of these models are consistent, at least qualitatively, with the results on optical luminosity dependence of the quasar clustering, obtained by different authors \citep{croom_2005,porciani_2006,myers_2006,myers_2007_2,daAngela_2008,padmanabhan+2009,Shen_2009,white+2012}, and in the present work. On the other hand, the result clustering of X-ray selected quasars and other types of AGNs require more detailed models of AGN formation (e.\,g. \citealt{croton+2006,thacker+2008}). For discussion of this problem, see e.\,g. \citealt{coil+2009,cappelluti+2012}. One of the main consequence of the smaller bias of the optically selected quasars than that of the X-ray selected AGNs, is that the theories of the quasar formation, in which AGNs are fuelled only by major mergers is not enough to describe these results and other mechanisms like tidal disruption or disk instability are required to be additional triggers of AGN activity.

\section{Conclusions}\label{sec:6}

\indent\indent Using the SDSS NBCKDE catalogue of photometrically selected quasar candidates \citep{richards_2009} drawn from the 6th release of the SDSS we studied the angular and spatial clustering of quasars with the help of the 2pCF. For this purpose own technique of the random catalogue generation was investigated and used. The catalogue is 3 times larger than that used in previous studies by \citet{myers_2006,myers_2007,myers_2007_2}, which allowed us to place more tight restrictions on the samples, namely to reject objects with photometric redshift determination probability less than 0.5, and to construct the main sample only from objects with redshifts within the `SDSS window', $0.8<z<2.2$ \citep{weinstein_2004}. Hence, even having the main sample with 230\,829 objects, which is not larger than the samples used by Myers at al., our results should be viewed as a corroboration of these previous works, with the main utilities being our use of an independent technique to generate random catalogues, and 
also use of a sample with ``higher quality''. The main results of this work are the following:
\begin{enumerate}
 \item The angular 2pCF of photometrically selected quasars, averaged over $0.8<z_{phot}<2.2$, is approximated well with the power law $w(\theta)=\left(\theta/\theta_{0}\right)^{-\alpha}$ with $\theta_{0}=4''.5\pm1''.4$ and $\alpha=0.94\pm0.06$ over the range $1'<\theta<40'$, that agree well with previous results by \citet{myers_2006,myers_2007} for similar mean redshift, but averaged over broader $z_{phot}$ range.
 \item The parameters of the deprojected 2pCF averaged over $0.8<z_{phot}<2.2$ and modelled with a power law $\xi(r)=\left(r/r_{0}\right)^{-\gamma}$, are $r_{0}=7.81^{+1.18}_{-1.16}\,h^{-1}$\,Mpc, $\gamma=1.94\pm0.06$, which are in perfect agreement with previous results from spectroscopic surveys. 
 \item Splitting the sample into three redshift ranges, we confirmed the evidence for an increase of the clustering amplitude with redshift, found in previous studies of photometrically selected and spectroscopically confirmed quasar samples.  \item Selecting three subsamples with different $M_{i}$ and $0.8<z_{phot}<2.2$, we have found no evidence for luminosity dependence of the quasar clustering. The marginal decrease in correlation length for the sample of \textit{faint} quasars can be addressed to the redshift evolution, which is hard to avoid when analysing different luminosity subsamples within a broad redshift range. This result is consistent with the models of the quasar formation, in which bright and faint quasars are assumed to be similar sources, hosted by dark matter halos of similar masses, but observed at different stages of their evolution.
 \item Comparison of our results with studies of the X-ray selected AGNs with similar redshift showed that the clustering amplitude of optically selected quasars is similar to that of X-ray selected quasars, but it is lower than that of samples of all X-ray selected AGNs. As far as X-ray selected AGN samples contain both AGN types, this result evidences for different types of AGNs to reside in different environments.
 \item We also investigated our sample and found that in addition to the well-known stellar contamination of photometrically selected quasar candidates there is also a small (about 0.1\%) contamination by artifacts of the automatic selection technique of point sources. These include star formation regions in spiral galaxies and diffraction spikes around bright stars. These artifacts were already extensively discussed in \citet{myers_2007_2}. The subsamples with high and low Galactic extinction (reddening) showed similar results on 2pCF, suggesting that Galactic reddening does not strongly influence quasar clustering when our technique for random-catalogue-generation is adopted.
\end{enumerate}

\vspace*{-2ex}
\section*{Acknowledgements}

\indent\indent The authors are thankful to V.\,Ya.\,Choliy for his invaluable advices concerning development of the software for data treatment, and to V.\,I.\,Zhdanov, O.\,Ruchaysky and A.\,Boyarsky for useful comments and fruitful discussions. We are particularly indebted to the referee Adam Myers and another anonymous referee, who provided helpful and constructive criticism of the early version of the present paper.

The authors are also thankful to the Sloan Digital Sky Survey team. Funding for the SDSS and SDSS-II has been provided by the Alfred P.\,Sloan Foundation, the Participating Institutions, the National Science Foundation, the U.S. Department of Energy, the National Aeronautics and Space Administration, the Japanese Monbukagakusho, the Max Planck Society, and the Higher Education Funding Council for England. The SDSS Web Site is http://www.sdss.org/. The SDSS is managed by the Astrophysical Research Consortium for the Participating Institutions. The Participating Institutions are the American Museum of Natural History, Astrophysical Institute Potsdam, University of Basel, University of Cambridge, Case Western Reserve University, University of Chicago, Drexel University, Fermilab, the Institute for Advanced Study, the Japan Participation Group, Johns Hopkins University, the Joint Institute for Nuclear Astrophysics, the Kavli Institute for Particle Astrophysics and Cosmology, the Korean Scientist Group, the 
Chinese Academy of Sciences (LAMOST), Los Alamos National Laboratory, the Max-Planck-Institute for Astronomy (MPIA), the Max-Planck-Institute for Astrophysics (MPA), New Mexico State University, Ohio State University, University of Pittsburgh, University of Portsmouth, Princeton University, the United States Naval Observatory, and the University of Washington.

This work has been  partially supported by Swiss National Science Foundation  (SCOPES grant No 128040). 

\vspace*{-2ex}

\begin{thebibliography}{}

\bibitem[\protect\citeauthoryear{{Abazajian} \& {et al.}}{{Abazajian} et~al.}{2009}]{Abazadjian_2009}
{Abazajian} K.~N.  {et al.}, 2009, \apjs, 182, 543

\bibitem[\protect\citeauthoryear{{Allevato} et~al.}{2011}]{Allevato+2011}
{Allevato} V.  {et al.},  2011, \apj, 736, 99

\bibitem[\protect\citeauthoryear{{Bardeen}, {Bond}, {Kaiser} \&
  {Szalay}}{{Bardeen} et~al.}{1986}]{bardeen+1986}
{Bardeen} J.~M.,  {Bond} J.~R.,  {Kaiser} N.,    {Szalay} A.~S.,  1986, \apj,
  304, 15

\bibitem[\protect\citeauthoryear{{Basilakos}, {Plionis}, {Georgakakis} \&
  {Georgantopoulos}}{{Basilakos} et~al.}{2005}]{Basilakos+2005}
{Basilakos} S.,  {Plionis} M.,  {Georgakakis} A.,    {Georgantopoulos} I.,
  2005, \mnras, 356, 183

\bibitem[\protect\citeauthoryear{{Beisbart} \& {Kerscher}}{{Beisbart} \&
  {Kerscher}}{2000}]{beisbart_2000}
{Beisbart} C.,  {Kerscher} M.,  2000, \apj, 545, 6

\bibitem[\protect\citeauthoryear{{Brunner}, {Szalay} \& {Connolly}}{{Brunner}
  et~al.}{2000}]{brunner+2000}
{Brunner} R.~J.,  {Szalay} A.~S.,    {Connolly} A.~J.,  2000, \apj, 541, 527

\bibitem[\protect\citeauthoryear{{Cappelluti}, {Allevato} \&
  {Finoguenov}}{{Cappelluti} et~al.}{2012}]{cappelluti+2012}
{Cappelluti} N.,  {Allevato} V.,    {Finoguenov} A.,  2012, Advances in
  Astronomy, 2012

\bibitem[\protect\citeauthoryear{{Carroll}, {Press} \& {Turner}}{{Carroll}
  et~al.}{1992}]{carroll+1992}
{Carroll} S.~M.,  {Press} W.~H.,    {Turner} E.~L.,  1992, \araa, 30, 499

\bibitem[\protect\citeauthoryear{{Coil} et~al.}{2009}]{coil+2009}
{Coil} A.~L.  {et al.},  2009, \apj, 701, 1484

\bibitem[\protect\citeauthoryear{{Coil}, {Hennawi}, {Newman}, {Cooper} \&
  {Davis}}{{Coil} et~al.}{2007}]{coil_2007}
{Coil} A.~L.,  {Hennawi} J.~F.,  {Newman} J.~A.,  {Cooper} M.~C.,    {Davis}
  M.,  2007, \apj, 654, 115

\bibitem[\protect\citeauthoryear{{Coil} et al.}{{Coil}
  et~al.}{2008}]{coil_2008}{Coil} A.~L.  {et al.},  2008, \apj, 672, 153

\bibitem[\protect\citeauthoryear{{Cole} \& {Kaiser}}{{Cole} \&
  {Kaiser}}{1989}]{cole+1989}
{Cole} S.,  {Kaiser} N.,  1989, \mnras, 237, 1127

\bibitem[\protect\citeauthoryear{{Croom}, {Boyle}, {Loaring}, {Miller},
  {Outram}, {Shanks} \& {Smith}}{{Croom} et~al.}{2002}]{Croom+2002}
{Croom} S.~M.,  {Boyle} B.~J.,  {Loaring} N.~S.,  {Miller} L.,  {Outram} P.~J.,
   {Shanks} T.,    {Smith} R.~J.,  2002, \mnras, 335, 459

\bibitem[\protect\citeauthoryear{{Croom} et al.}{{Croom}
  et~al.}{2005}]{croom_2005} {Croom} S.~M.  {et al.},  2005,
  \mnras, 356, 415

\bibitem[\protect\citeauthoryear{{Croom} \& {Shanks}}{{Croom} \&
  {Shanks}}{1996}]{croom_shanks_1996}
{Croom} S.~M.,  {Shanks} T.,  1996, \mnras, 281, 893

\bibitem[\protect\citeauthoryear{{Croom}, {Smith}, {Boyle}, {Shanks}, {Miller},
  {Outram} \& {Loaring}}{{Croom} et~al.}{2004}]{Croom_2004}
{Croom} S.~M.,  {Smith} R.~J.,  {Boyle} B.~J.,  {Shanks} T.,  {Miller} L.,
  {Outram} P.~J.,    {Loaring} N.~S.,  2004, \mnras, 349, 1397

\bibitem[\protect\citeauthoryear{{Croton} et al.}{{Croton}
  et~al.}{2006}]{croton+2006}
{Croton} D.~J.  {et al.},  2006, \mnras, 365, 11

\bibitem[\protect\citeauthoryear{{da {\^A}ngela}, {Outram}, {Shanks}, {Boyle},
  {Croom}, {Loaring}, {Miller} \& {Smith}}{{da {\^A}ngela}
  et~al.}{2005}]{daAngela_2005}
{da {\^A}ngela} J.,  {Outram} P.~J.,  {Shanks} T.,  {Boyle} B.~J.,  {Croom}
  S.~M.,  {Loaring} N.~S.,  {Miller} L.,    {Smith} R.~J.,  2005, \mnras, 360,
  1040

\bibitem[\protect\citeauthoryear{{da {\^A}ngela} et al.}{{da {\^A}ngela}
  et~al.}{2008}]{daAngela_2008} {da {\^A}ngela} J.  {et al.},  2008, \mnras, 383, 565

\bibitem[\protect\citeauthoryear{{Ebrero} et al.}{{Ebrero}
  et~al.}{2009}]{ebrero_2009_a}{Ebrero} J.  {et al.},
  2009, \aap, 493, 55

\bibitem[\protect\citeauthoryear{{Ebrero}, {Mateos}, {Stewart}, {Carrera} \&
  {Watson}}{{Ebrero} et~al.}{2009}]{ebrero_2009}
{Ebrero} J.,  {Mateos} S.,  {Stewart} G.~C.,  {Carrera} F.~J.,    {Watson}
  M.~G.,  2009, \aap, 500, 749

\bibitem[\protect\citeauthoryear{{Einasto} et~al.}{2007}]{einasto_2007}
{Einasto} M.  {et al.},  2007, preprint (arXiv:0706.1126)

\bibitem[\protect\citeauthoryear{{Elyiv} et~al.}{2012}]{elyiv_2011}
{Elyiv} A.  {et al.},  2012, \aap, 537, A131

\bibitem[\protect\citeauthoryear{{Ferrarese}}{{Ferrarese}}{2002}]{Ferrarese_20%
02}{Ferrarese} L.,  2002, \apj, 578, 90

\bibitem[\protect\citeauthoryear{{Gilli} et~al.}{2005}]{Gilli+2005}
{Gilli} R.  {et al.},  2005, \aap, 430, 811

\bibitem[\protect\citeauthoryear{{Gilli} et~al.}{2009}]{Gilli+2009}
{Gilli} R.  {et al.},  2009, \aap, 494, 33

\bibitem[\protect\citeauthoryear{{Haiman} \& {Hui}}{{Haiman} \&
  {Hui}}{2001}]{haiman+01}
{Haiman} Z.,  {Hui} L.,  2001, \apj, 547, 27

\bibitem[\protect\citeauthoryear{{Hamilton}, {Kumar}, {Lu} \&
  {Matthews}}{{Hamilton} et~al.}{1991}]{hamilton_1991}
{Hamilton} A.~J.~S.,  {Kumar} P.,  {Lu} E.,    {Matthews} A.,  1991, \apjl,
  374, L1

\bibitem[\protect\citeauthoryear{{Hamilton}, {Kumar}, {Lu} \&
  {Matthews}}{{Hamilton} et~al.}{1995}]{hamilton_1995}
{Hamilton} A.~J.~S.,  {Kumar} P.,  {Lu} E.,    {Matthews} A.,  1995, \apjl,
  442, L73

\bibitem[\protect\citeauthoryear{{Hennawi}  {et al.}}{{Hennawi} et~al.}{2006}]{hennawi_2006}
{Hennawi} J.~F.  {et al.},  2006, \aj, 131, 1

\bibitem[\protect\citeauthoryear{{Hopkins}, {Hernquist}, {Cox}, {Di Matteo},
  {Robertson} \& {Springel}}{{Hopkins} et~al.}{2005a}]{hopkins+2005}
{Hopkins} P.~F.,  {Hernquist} L.,  {Cox} T.~J.,  {Di Matteo} T.,  {Robertson}
  B.,    {Springel} V.,  2005a, \apj, 630, 716

\bibitem[\protect\citeauthoryear{{Hopkins}, {Hernquist}, {Cox}, {Di Matteo},
  {Robertson} \& {Springel}}{{Hopkins} et~al.}{2005b}]{hopkins+2005_2}
{Hopkins} P.~F.,  {Hernquist} L.,  {Cox} T.~J.,  {Di Matteo} T.,  {Robertson}
  B.,    {Springel} V.,  2005b, \apj, 632, 81

\bibitem[\protect\citeauthoryear{{Hopkins}, {Hernquist}, {Cox} \& {Kere{\v
  s}}}{{Hopkins} et~al.}{2008}]{hopkins+2008}
{Hopkins} P.~F.,  {Hernquist} L.,  {Cox} T.~J.,    {Kere{\v s}} D.,  2008,
  \apjs, 175, 356
  
\bibitem[\protect\citeauthoryear{{Ivashchenko}}{{Ivashchenko}}{2008}]{ivashche%
nko_2008_lum}
{Ivashchenko} G.,  2008, Journal of Physical Studies, 12, 3902  

\bibitem[\protect\citeauthoryear{{Ivashchenko}, {Zhdanov} \&
  {Tugay}}{{Ivashchenko} et~al.}{2010}]{ivashchenko_2010}
{Ivashchenko} G.,  {Zhdanov} V.~I.,    {Tugay} A.~V.,  2010, \mnras, 409, 1691

\bibitem[\protect\citeauthoryear{{Ivashchenko} \& {Zhdanov}}{{Ivashchenko} \&
  {Zhdanov}}{2010}]{ivashchenko_2010_dr5}
{Ivashchenko} G.~Y.,  {Zhdanov} V.~I.,  2010, Kinematika i Fizika Nebesnykh
  Tel, 26, 43

\bibitem[\protect\citeauthoryear{{Jenkins}  {et al.}}{{Jenkins} et~al.}{1998}]{jenkins+1998}
{Jenkins} A.  {et al.},  1998, \apj, 499, 20

\bibitem[\protect\citeauthoryear{{Kaiser}}{{Kaiser}}{1984}]{kaiser_1984}
{Kaiser} N.,  1984, \apjl, 284, L9

\bibitem[\protect\citeauthoryear{{Kaiser}}{{Kaiser}}{1987}]{kaiser_1987}
{Kaiser} N.,  1987, \mnras, 227, 1

\bibitem[\protect\citeauthoryear{{Kauffmann} \& {Haehnelt}}{{Kauffmann} \&
  {Haehnelt}}{2000}]{kauffmann+2000}
{Kauffmann} G.,  {Haehnelt} M.,  2000, \mnras, 311, 576

\bibitem[\protect\citeauthoryear{{Krumpe}, {Miyaji} \& {Coil}}{{Krumpe}
  et~al.}{2010}]{krumpe+2010}
{Krumpe} M.,  {Miyaji} T.,    {Coil} A.~L.,  2010, \apj, 713, 558

\bibitem[\protect\citeauthoryear{{Landy} \& {Szalay}}{{Landy} \&
  {Szalay}}{1993}]{landy_szalay_1993}
{Landy} S.~D.,  {Szalay} A.~S.,  1993, \apj, 412, 64

\bibitem[\protect\citeauthoryear{{Lidz}, {Hopkins}, {Cox}, {Hernquist} \&
  {Robertson}}{{Lidz} et~al.}{2006}]{lidz+2006}
{Lidz} A.,  {Hopkins} P.~F.,  {Cox} T.~J.,  {Hernquist} L.,    {Robertson} B.,
  2006, \apj, 641, 41

\bibitem[\protect\citeauthoryear{{Limber}}{{Limber}}{1953}]{limber_1953}
{Limber} D.~N.,  1953, \apj, 117, 134

\bibitem[\protect\citeauthoryear{{Martini} \& {Weinberg}}{{Martini} \&
  {Weinberg}}{2001}]{martini+01}
{Martini} P.,  {Weinberg} D.~H.,  2001, \apj, 547, 12

\bibitem[\protect\citeauthoryear{{Miyaji}, {Hasinger} \& {Schmidt}}{{Miyaji}
  et~al.}{2000}]{Miyaji+2000}
{Miyaji} T.,  {Hasinger} G.,    {Schmidt} M.,  2000, \aap, 353, 25

\bibitem[\protect\citeauthoryear{{Miyaji} et~al.}{2007}]{Miyaji+2007}
{Miyaji} T.  {et al.},  2007, \apjs, 172, 396

\bibitem[\protect\citeauthoryear{{Mo} \& {White}}{{Mo} \&
  {White}}{1996}]{mo+1996}
{Mo} H.~J.,  {White} S.~D.~M.,  1996, \mnras, 282, 347

\bibitem[\protect\citeauthoryear{{Mountrichas}, {Sawangwit}, {Shanks}, {Croom},
  {Schneider}, {Myers} \& {Pimbblet}}{{Mountrichas}
  et~al.}{2009}]{Mountrichas_2009}
{Mountrichas} G.,  {Sawangwit} U.,  {Shanks} T.,  {Croom} S.~M.,  {Schneider}
  D.~P.,  {Myers} A.~D.,    {Pimbblet} K.,  2009, \mnras, 394, 2050

\bibitem[\protect\citeauthoryear{{Myers}, {Brunner}, {Nichol}, {Richards},
  {Schneider} \& {Bahcall}}{{Myers} et~al.}{2007a}]{myers_2007}
{Myers} A.~D.,  {Brunner} R.~J.,  {Nichol} R.~C.,  {Richards} G.~T.,
  {Schneider} D.~P.,    {Bahcall} N.~A.,  2007, \apj, 658, 85
  
\bibitem[\protect\citeauthoryear{{Myers}, {Brunner}, {Richards}, {Nichol}, {Schneider}, \& {Bahcall}}{{Myers} et~al.}{2007b}]{myers_2007_2}
{Myers}, A.~D., {Brunner}, R.~J., {Richards}, G.~T., {Nichol}, R.~C., {Schneider}, D.~P., {Bahcall}, N.~A.,  2007, \apj, 658, 85
  
\bibitem[\protect\citeauthoryear{{Myers} et~al.}{2006}]{myers_2006}
{Myers} A.~D.  {et al.},  2006, \apj, 638, 622

\bibitem[\protect\citeauthoryear{{Owers}, {Blake}, {Couch}, {Pracy} \&
  {Bekki}}{{Owers} et~al.}{2007}]{owners_2007}
{Owers} M.~S.,  {Blake} C.,  {Couch} W.~J.,  {Pracy} M.~B.,    {Bekki} K.,
  2007, \mnras, 381, 494

\bibitem[\protect\citeauthoryear{{Padmanabhan}, {White}, {Norberg} \&
  {Porciani}}{{Padmanabhan} et~al.}{2009}]{padmanabhan+2009}
{Padmanabhan} N.,  {White} M.,  {Norberg} P.,    {Porciani} C.,  2009, \mnras,
  397, 1862

\bibitem[\protect\citeauthoryear{{Peebles}}{{Peebles}}{1980}]{peebles_book}
{Peebles} P.~J.~E.,  1980, {The large-scale structure of the universe}

\bibitem[\protect\citeauthoryear{{Porciani}, {Magliocchetti} \&
  {Norberg}}{{Porciani} et~al.}{2004}]{porciani_2004}
{Porciani} C.,  {Magliocchetti} M.,    {Norberg} P.,  2004, \mnras, 355, 1010

\bibitem[\protect\citeauthoryear{{Porciani} \& {Norberg}}{{Porciani} \&
  {Norberg}}{2006}]{porciani_2006}
{Porciani} C.,  {Norberg} P.,  2006, \mnras, 371, 1824

\bibitem[\protect\citeauthoryear{{Richards} et~al.}{2005}]{richards_2005}
{Richards} G.~T.  {et al.},  2005, \mnras, 360, 839

\bibitem[\protect\citeauthoryear{{Richards} {et al.}}{{Richards}
  et~al.}{2009}]{richards_2009}{Richards} G.~T.  {et al.},  2009, \apjs, 180, 67

\bibitem[\protect\citeauthoryear{{Richards}  {et al.}}{{Richards} et~al.}{2004}]{richards_2004}
{Richards} G.~T.  {et al.},  2004, \apjs, 155, 257

\bibitem[\protect\citeauthoryear{{Richards} et~al.}{2006}]{richards_2006}
{Richards} G.~T.  {et al.},  2006, \aj, 131, 2766

\bibitem[\protect\citeauthoryear{{Ross}, {Brunner} \& {Myers}}{{Ross}
  et~al.}{2007}]{ross_2007}
{Ross} A.~J.,  {Brunner} R.~J.,    {Myers} A.~D.,  2007, \apj, 665, 67

\bibitem[\protect\citeauthoryear{{Ross} et~al.}{2009}]{ross_2009}
{Ross} N.~P.  {et al.},  2009, \apj, 697, 1634

\bibitem[\protect\citeauthoryear{{Schlegel}, {Finkbeiner} \&
  {Davis}}{{Schlegel} et~al.}{1998}]{schlegel_1998}
{Schlegel} D.~J.,  {Finkbeiner} D.~P.,    {Davis} M.,  1998, \apj, 500, 525

\bibitem[\protect\citeauthoryear{{Schneider} et~al.}{2007}]{schneider+2007}
{Schneider} D.~P.  {et al.},  2007, \aj, 134, 102

\bibitem[\protect\citeauthoryear{{Schneider} et al.}{{Schneider} et~al.}{2005}]{schneider+2005}{Schneider} D.~P. {et al.},  2005, \aj, 130, 367

\bibitem[\protect\citeauthoryear{{SDSS-III Collaboration: Christopher P.~Ahn} et al.}{{SDSS-III Collaboration: Christopher
  P.~Ahn} et~al.}{2012}]{dr9}
SDSS-III Collaboration: Christopher P.~Ahn  {et al.}, 2012, preprint (arXiv:1207.7137)

\bibitem[\protect\citeauthoryear{{Shanks} \& {Boyle}}{{Shanks} \&
  {Boyle}}{1994}]{shanks_boyle_1994}
{Shanks} T.,  {Boyle} B.~J.,  1994, \mnras, 271, 753

\bibitem[\protect\citeauthoryear{{Shen} et~al.}{2010}]{Shen_II_2010}
{Shen} Y. {et al.},  2010, \apj, 719, 1693

\bibitem[\protect\citeauthoryear{{Shen} {et al.}}{{Shen}
  et~al.}{2007}]{shen_2007}
{Shen} Y. {et al.},  2007, \aj, 133, 2222

\bibitem[\protect\citeauthoryear{{Shen} {et al.}}{{Shen} et~al.}{2009}]{Shen_2009}
{Shen} Y. {et al.},  2009, \apj, 697, 1656

\bibitem[\protect\citeauthoryear{{Sheth} \& {Lemson}}{{Sheth} \&
  {Lemson}}{1999}]{sheth+1999}
{Sheth} R.~K.,  {Lemson} G.,  1999, \mnras, 304, 767

\bibitem[\protect\citeauthoryear{{Sheth}, {Mo} \& {Tormen}}{{Sheth}
  et~al.}{2001}]{Sheth+2001}
{Sheth} R.~K.,  {Mo} H.~J.,    {Tormen} G.,  2001, \mnras, 323, 1

\bibitem[\protect\citeauthoryear{{Smith} et~al.}{2003}]{smith_2003}
{Smith} R.~E. {et al.},  2003, \mnras, 341, 1311

\bibitem[\protect\citeauthoryear{{Sorrentino}, {Antonuccio-Delogu} \&
  {Rifatto}}{{Sorrentino} et~al.}{2006}]{sorrentino_2006}
{Sorrentino} G.,  {Antonuccio-Delogu} V.,    {Rifatto} A.,  2006, {\aa}p, 460,
  673

\bibitem[\protect\citeauthoryear{{Thacker}, {Scannapieco}, {Couchman} \&
  {Richardson}}{{Thacker} et~al.}{2009}]{thacker+2008}
{Thacker} R.~J.,  {Scannapieco} E.,  {Couchman} H.~M.~P.,    {Richardson} M.,
  2009, \apj, 693, 552

\bibitem[\protect\citeauthoryear{{Ueda}, {Akiyama}, {Ohta} \& {Miyaji}}{{Ueda}
  et~al.}{2003}]{Ueda+2003}
{Ueda} Y.,  {Akiyama} M.,  {Ohta} K.,    {Miyaji} T.,  2003, \apj, 598, 886

\bibitem[\protect\citeauthoryear{{Weinstein} et~al.}{2004}]{weinstein_2004}
{Weinstein} M.~A. {et~al.},  2004, \apjs, 155, 243

\bibitem[\protect\citeauthoryear{{White} {et al.}}{{White}
  et~al.}{2012}]{white+2012}{White} M. {et~al.},
  2012, \mnras, 424, 933

\bibitem[\protect\citeauthoryear{{Wyithe} \& {Loeb}}{{Wyithe} \&
  {Loeb}}{2003}]{wyithe+2003}
{Wyithe} J.~S.~B.,  {Loeb} A.,  2003, \apj, 595, 614

\bibitem[\protect\citeauthoryear{{Wyithe} \& {Loeb}}{{Wyithe} \&
  {Loeb}}{2005}]{Wyithe+2005}
{Wyithe} J.~S.~B.,  {Loeb} A.,  2005, \apj, 621, 95

\bibitem[\protect\citeauthoryear{{Yang}, {Mushotzky}, {Barger} \&
  {Cowie}}{{Yang} et~al.}{2006}]{Yang+2006}
{Yang} Y.,  {Mushotzky} R.~F.,  {Barger} A.~J.,    {Cowie} L.~L.,  2006, \apj,
  645, 68

\bibitem[\protect\citeauthoryear{{Zehavi} et~al.}{2005}]{zehavi_2005}
{Zehavi} I. {et~al.},  2005, \apj, 630, 1

\bibitem[\protect\citeauthoryear{{Zhdanov} \& {Ivashchenko}}{{Zhdanov} \&
  {Ivashchenko}}{2008}]{ivashchenko_2008_dr3}
{Zhdanov} V.~I.,  {Ivashchenko} G.~Y.,  2008, Kinematika i Fizika Nebesnykh
  Tel, 24, 3

\end{thebibliography}
\let\jnlstyle=\rm\def\jref#1{{\jnlstyle#1}}\def\aj{\jref{AJ}}
  \def\araa{\jref{ARA\&A}} \def\apj{\jref{ApJ}} \def\apjl{\jref{ApJ}}
  \def\apjs{\jref{ApJS}} \def\ao{\jref{Appl.~Opt.}} \def\apss{\jref{Ap\&SS}}
  \def\aap{\jref{A\&A}} \def\aapr{\jref{A\&A~Rev.}} \def\aaps{\jref{A\&AS}}
  \def\azh{\jref{AZh}} \def\baas{\jref{BAAS}} \def\jrasc{\jref{JRASC}}
  \def\memras{\jref{MmRAS}} \def\mnras{\jref{MNRAS}}
  \def\pra{\jref{Phys.~Rev.~A}} \def\prb{\jref{Phys.~Rev.~B}}
  \def\prc{\jref{Phys.~Rev.~C}} \def\prd{\jref{Phys.~Rev.~D}}
  \def\pre{\jref{Phys.~Rev.~E}} \def\prl{\jref{Phys.~Rev.~Lett.}}
  \def\pasp{\jref{PASP}} \def\pasj{\jref{PASJ}} \def\qjras{\jref{QJRAS}}
  \def\skytel{\jref{S\&T}} \def\solphys{\jref{Sol.~Phys.}}
  \def\sovast{\jref{Soviet~Ast.}} \def\ssr{\jref{Space~Sci.~Rev.}}
  \def\zap{\jref{ZAp}} \def\nat{\jref{Nature}} \def\iaucirc{\jref{IAU~Circ.}}
  \def\aplett{\jref{Astrophys.~Lett.}}
  \def\apspr{\jref{Astrophys.~Space~Phys.~Res.}}
  \def\bain{\jref{Bull.~Astron.~Inst.~Netherlands}}
  \def\fcp{\jref{Fund.~Cosmic~Phys.}} \def\gca{\jref{Geochim.~Cosmochim.~Acta}}
  \def\grl{\jref{Geophys.~Res.~Lett.}} \def\jcp{\jref{J.~Chem.~Phys.}}
  \def\jgr{\jref{J.~Geophys.~Res.}}
  \def\jqsrt{\jref{J.~Quant.~Spec.~Radiat.~Transf.}}
  \def\memsai{\jref{Mem.~Soc.~Astron.~Italiana}}
  \def\nphysa{\jref{Nucl.~Phys.~A}} \def\physrep{\jref{Phys.~Rep.}}
  \def\physscr{\jref{Phys.~Scr}} \def\planss{\jref{Planet.~Space~Sci.}}
  \def\procspie{\jref{Proc.~SPIE}} \let\astap=\aap \let\apjlett=\apjl
  \let\apjsupp=\apjs \let\applopt=\ao

\label{lastpage}
\end{document}